\documentclass[aps,prb,twocolumn,superscriptaddress]{revtex4-2}
\usepackage{ucs}
\usepackage{natbib}
\usepackage{graphicx}  
\usepackage{bm}        
\usepackage{amssymb}   
\usepackage{amsmath}
\usepackage{color}
\usepackage{CJK}
\usepackage{times}
\usepackage{verbatim}
\usepackage{mathrsfs}
\usepackage{hyperref}
\usepackage{ulem}
\usepackage{physics}
\hypersetup{colorlinks = True, urlcolor=blue, linkcolor=blue, citecolor=blue}

\newcommand{\bg}{\begin{pmatrix}}
\newcommand{\ed}{\end{pmatrix}}

\usepackage{cancel}
\usepackage{comment}

\usepackage{footmisc}

\usepackage{graphicx}
\usepackage{dcolumn}
\usepackage{bm}
\usepackage[]{xspace}
\usepackage{hyperref}

\graphicspath{{./}{./Figs/}{./figs}{./psfigs/}{./Figures/}}
\usepackage{xcolor}

\begin{document}

\title{Spin-strain interactions under hydrostatic pressure in \texorpdfstring{$\alpha$-RuCl$_3$}{alpha-RuCl3}} 


\author{A. Hauspurg}
\affiliation{Hochfeld-Magnetlabor Dresden (HLD-EMFL) and W\"urzburg-Dresden Cluster of Excellence ct.qmat, Helmholtz-Zentrum Dresden-Rossendorf (HZDR), 01328 Dresden, Germany}
\affiliation{Institut f\"ur Festk\"orper- und Materialphysik, TU Dresden, 01062 Dresden, Germany}

\author{Susmita Singh}
\affiliation{School of Physics and Astronomy, University of Minnesota, Minneapolis, Minnesota 55455, USA}
\author{T. Yanagisawa}
\affiliation{Department of Physics, Hokkaido University, Sapporo 060-0810, Japan}

\author{V. Tsurkan}
\affiliation{Experimental Physics V, Center for Electronic Correlations and Magnetism, University of Augsburg, 86135 Augsburg, Germany}
\affiliation{Institute of Applied Physics, Moldova State University, MD 2028, Chisinau, Republic of Moldova}

\author{J. Wosnitza}
\affiliation{Hochfeld-Magnetlabor Dresden (HLD-EMFL) and W\"urzburg-Dresden Cluster of Excellence ct.qmat, Helmholtz-Zentrum Dresden-Rossendorf (HZDR), 01328 Dresden, Germany}
\affiliation{Institut f\"ur Festk\"orper- und Materialphysik, TU Dresden, 01062 Dresden, Germany}

\author{Wolfram Brenig}
\affiliation{Institute for Theoretical Physics, Technical University Braunschweig, 38106 Braunschweig, Germany}

\author{Natalia B. Perkins}
\affiliation{School of Physics and Astronomy, University of Minnesota, Minneapolis, Minnesota 55455, USA}
\affiliation{Technical University of Munich, Germany, Institute for Advanced Study, 85748 Garching, Germany}

\author{S. Zherlitsyn}
\affiliation{Hochfeld-Magnetlabor Dresden (HLD-EMFL) and W\"urzburg-Dresden Cluster of Excellence ct.qmat, Helmholtz-Zentrum Dresden-Rossendorf (HZDR), 01328 Dresden, Germany}


\date{\today}

\begin{abstract}
We investigate the effects of hydrostatic pressure on $\alpha$-RuCl$_3$, a prototypical material for the Kitaev spin model on a honeycomb lattice with a possible spin-liquid ground state. Using ultrasound measurements at pressures up to 1.16 GPa, we reveal significant modifications of the acoustic properties and the $H$-$T$ phase diagram of this material. Hydrostatic pressure suppresses the three-dimensional magnetic order and induces a dimerization transition at higher pressures. At low pressures, the sound attenuation exhibits a linear temperature dependence, while above 0.28 GPa, it becomes nearly temperature independent, suggesting a shift in the phonon scattering regime dominated by Majorana fermions. These findings provide new insights into spin-strain interactions in Kitaev magnets and deliver a detailed characterization of the $H$-$T$ phase diagram of $\alpha$-RuCl$_3$ under hydrostatic pressure.
\end{abstract}

\date{\today}
\maketitle





\section{Introduction}
In recent years, significant efforts have been focused on finding materials with bond-dependent Ising-like interactions that could realize the spin-1/2 Kitaev honeycomb model \cite{jackeli09,takagiconcept2019,Trebst2022,Rousochatzakis2024} and its quantum spin-liquid (QSL) state with fractionalized excitations -- Majorana fermions and $Z_{2}$ fluxes \cite{kitaevanyons2006}. Although, so far, no material has been confirmed as a perfect realization of this idealized model,
 the $J_{\textit{eff}}$ = $1/2$  strongly spin-orbit-coupled Mott insulator $\alpha$-RuCl$_3$  \cite{Plumb2014,Sears2015,kasaharamajorana2018,tanakathermodynamic2022,yokoihalfinteger2021,balzfield2021,suzukiproximate2021,wagnermagnetooptical2022,bachusthermodynamic2020} stands out as a promising candidate to investigate the intriguing physics of Kitaev quantum spin liquids.
Despite exhibiting a zigzag antiferromagnetic (AFM) order below 
 7 K, indicating deviations from the pure Kitaev model due to subdominant non-Kitaev interactions,  various dynamical probes suggest that  $\alpha$-RuCl$_3$  lies in close proximity to a QSL state
  \cite{ banerjeeneutron2017,kasaharamajorana2018,tanakathermodynamic2022,yokoihalfinteger2021,wolterspin2022}.  
 Among the many dynamical probes proposed to study this proximity 
~\cite{knolle14,knolle2014raman,perreault2015theory,perreault2016resonant,rousochatzakis2019quantum,Gabor2016,halasz2019observing,wan2019resolving,pereira2020electrical,udagawa2021scanning,joy2022dynamics,bauer2023scanning,kao2024dynamics,kao2024STM}, 
recent studies have identified phonon dynamics as a particularly promising and insightful tool. 
This approach is particularly relevant for $\alpha$-RuCl$_3$, where spin-lattice coupling plays a crucial role in its various properties and phenomena \cite{kasaharamajorana2018, aviv18,ye18, metavitsiadisphonon2020, yephonon2020, ligiant2021,kocsismagnetoelastic2022, kaibmagnetoelastic2021, schoenemannthermal2020,hentrichunusual2018,   reschketerahertz2019, bruinrobustness2022, yamashitasample2020}. 

Ultrasound measurements of sound velocity and attenuation provide an effective method for investigating the dynamics of acoustic phonons \cite{truellultrasonic1969,luthiphysical2005}, a technique recently applied to  $\alpha$-RuCl$_3$ to explore its temperature and magnetic-field dependences of sound attenuation \cite{hau24}.
This study revealed that the dominant processes contributing to the sound attenuation in this material can be attributed to phonons scattering off fractionalized excitations, thus supporting the idea  of proximity of $\alpha$-RuCl$_3$ to the Kitaev  QSL \cite{yephonon2020, metavitsiadisphonon2020, fengtemperature2021, sin23, Singh2024, Dantas2024}.

Lattice degrees of freedom have not only been of interest in the context of dynamical phonons coupled to a proximate QSL, but also significant attention has been devoted to studying the effects of hydrostatic pressure on $\alpha$-RuCl$_3$. In this work, we, therefore, make the natural step forward to combine the static and dynamic aspects of the coupling between the lattice, the magnetic, and electronic degrees of freedom in $\alpha$-RuCl$_3$, by investigating ultrasound propagation under hydrostatic pressure. This offers valuable insights into the nature of the magnetic state and the interplay between spin and lattice degrees of freedom.

A prime ingredient of our study is that $\alpha$-RuCl$_3$ under hydrostatic pressure undergoes a dimerization and a suppression of the magnetic ground state above a critical pressure $P_c$. To apply hydrostatic pressure, different experimental approaches have been utilized. The first involves the use of piston-type pressure cells, combined with pressure media such as Daphne 7373 \cite{cui17,bie18,bas18}. This is also employed in the present study.
 The second approach employs He-gas pressure cells \cite{wan23, wolf22, bas18, stanco24}. These cells offer the advantage of in situ pressure adjustment at low temperatures, provided the temperature remains above the crystallization point of $^4$He \cite{danielssimple1983}.
The third  approach by Wang \textit{et al.} in \cite{wangprb18} resorts to a toroid-type pressure cell with a glycerin/water mixture as the pressure medium, as well as a diamond anvil cell using NaCl as the pressure medium.

Remarkably, and while all methods agree on the existence of a critical pressure $P_c$, its reported value varies among the techniques. Using $^4$He as pressure medium, $P_c$ is 0.1–0.2 GPa \cite{cui17,bie18,bas18,stanco24}, whereas using Daphne 7373, $P_c$ is 0.7–1 GPa \cite{wan23,wolf22,bas18}.
A key observation is that pressure affects both spin and structural properties in a non-monotonic manner. Even before lattice dimerization, pressure partly suppresses magnetic ordering, evident as a gradual reduction in the Néel temperature, $T_N$, up to 0.25 GPa in piston-cell experiments \cite{wan23,wolf22}. Beyond 0.25 GPa, $T_N$ increases, leading to complete suppression of magnetic ordering at $P_c$.

To further explore the influence of such hydrostatic pressure on Kitaev physics and magnetoelastic coupling, we present a study employing ultrasound techniques, which are highly sensitive to phase transitions \cite{hau24, luthiphysical2005, truellultrasonic1969}, and examine how spin-strain interactions evolve under varying pressure. We observe that applied pressure increases the velocity of the transverse acoustic mode $(c_{11}-c_{12})/2$. However, this change is  nonlinear with pressure. Additionally, the attenuation of this mode changes from a linear temperature dependence to being nearly temperature independent.

To interpret these findings, we performed theoretical estimates of the sound attenuation. Our study reveals that the temperature dependence of the sound attenuation in $\alpha$-RuCl$_3$ is consistent with scattering of phonons by fractionalized excitations, a hallmark of the material's proximity to a Kitaev QSL. Our results also show that pressure-induced variations in the sound velocity and coupling constants are key to understanding the change of the sound attenuation versus pressure. Namely, within the pure Kitaev model, increasing pressure reduces the magnitude of the Kitaev coupling $K$ \cite{wolf22}. As our experiment shows an increase in sound velocity, this leads to a critical pressure, at which the sound velocity surpasses the Fermi velocity. This, in turn, causes the dominant scattering mechanism to shift from particle-hole to particle-particle processes, resulting in a transition from a linear increase of the sound attenuation with temperature, via an almost temperature-independent behavior, to a decrease with temperature \cite{yephonon2020, metavitsiadisphonon2020}.
Expanding our analysis to the $J$-$K$-$\Gamma$ model reveals similar trends, though the transition occurs at slightly higher pressures. This shift is attributed to the growing influence of Heisenberg $J$ and off-diagonal $\Gamma$ interactions, which further modify the Majorana fermion spectrum in $\alpha$-RuCl$_3$. 

We also investigated the pressure dependence of the $H$-$T$ phase diagram for $\alpha$-RuCl$_3$ with an in-plane magnetic field ($H \parallel a$), extending our measurements to hydrostatic pressures as high as 1.16 GPa. This approach provides valuable insights into the intricate coupling mechanisms and pressure-induced modifications of the magnetic and structural properties of this material.

\begin{figure}[tb]
	\centering
	\includegraphics[]{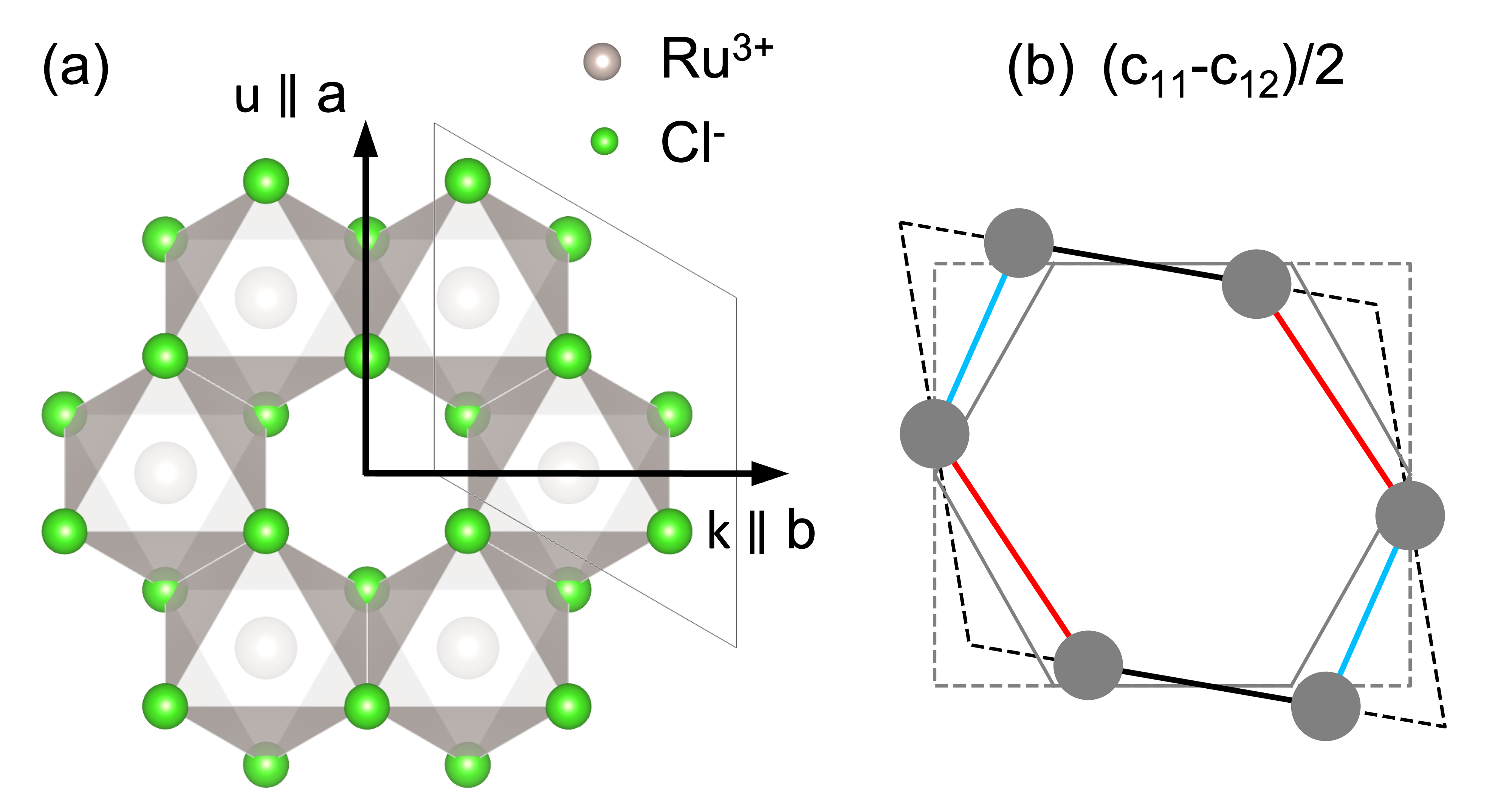}
	\caption{(a) Crystal structure of $\alpha$-RuCl$_3$ in a honeycomb plane and the crystallographic directions
	with the notations used in this work. Green and gray spheres indicate the positions of Cl$^-$ and Ru$^{3+}$ ions, respectively. The sound-propagation direction $k$ and polarization $u$ for the transverse acoustic mode $(c_{11}-c_{12})/2$ are shown. The thin lines represent the primitive unit cell in the $ab$ plane. Crystal structure created using Vesta \cite{mommavesta32011}. (b) Lattice strain (bold lines) related to the transverse acoustic mode $(c_{11}-c_{12})/2$ for a hexagonal lattice. Dashed lines represent the local symmetry before and after the deformation.}.
	\label{fig:CrystalStructure}
\end{figure}

\begin{figure}[tb]
	\centering
	\includegraphics[]{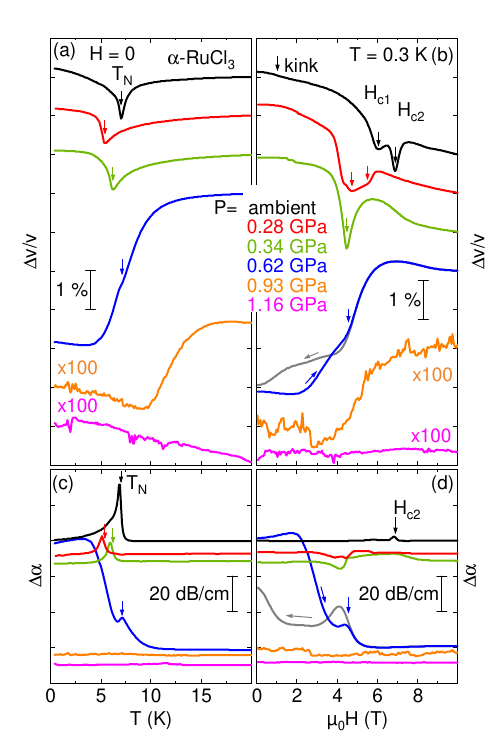}
	\caption{(a), (b) Relative change of the sound velocity and (c), (d) attenuation of the transverse acoustic mode $(c_{11}-c_{12})/2$  ($k \parallel b$, $u \parallel a$) versus temperature at $H = 0$ (a), (c) and versus magnetic field ($H \parallel a$) at 0.3 K (b), (d), measured at selected hydrostatic pressure values in $\alpha$-RuCl$_3$. The arrows indicate the critical temperatures and magnetic fields. The curves are arbitrarily shifted along the $y$ axis for clarity. Relative changes in $v$ and $\alpha$ are depicted by the corresponding scaling bar. Note, that the data at 0.93 and 1.16 GPa in panels (a) and (b) are scaled by a factor of 100. 
		\label{fig:Tdep_Hdep}}
\end{figure}

\section{Experimental details}
\subsection{Sample preparation and characterization}

We grew high-quality single crystals of $\alpha$-RuCl$_3$ by vacuum sublimation \cite{reschkesubgap2018,bachusthermodynamic2020}. 
The samples show a single ordering temperature of approximately 7 K [Fig. \ref{fig:Tdep_Hdep}(a)] and no signature of any additional phase transition at 14 K, typically caused by stacking faults \cite{mi21}. 
At room temperature, $\alpha$-RuCl$_3$ has a trigonal structure with $P$3$_1$12 space group \cite{banerjeeneutron2017,caolowtemperature2016}. 
Upon cooling, the material undergoes a first-order structural transition at about 140 K accompanied by a large hysteresis, though the precise low-temperature structure remains under debate \cite{murole2022, lebertacoustic2022}.
Weak van der Waals interactions between the layers may result in different stacking orders of the honeycomb planes, with minimal energy differences between stacking configurations \cite{yamauchilocal2018,caolowtemperature2016}. 

The samples are oriented using Laue x-ray backscattering diffraction. For clarity, we distinguish between the crystallographic directions within the honeycomb plane: The direction perpendicular to the Ru-Ru bonds (equivalent to $a$) and the direction parallel to the Ru-Ru bonds (equivalent to $b$), as illustrated in Fig. \ref{fig:CrystalStructure}(a).
For the ultrasound experiments the surfaces of the samples are treated with a focused ion beam (FIB) \cite{hau24}. 
This technique proves to process the sample surfaces with low mechanical wear, such that the fragile stacking order of the honeycomb planes is not disturbed. 
The sample dimensions are 1.63 mm along the sound propagation direction $b$, and approximately 0.5 mm along the $c$-axis, perpendicular to the honeycomb plane.

\subsection{Ultrasound measurement}

We performed ultrasound measurements using a pulse-echo phase-sensitive detection technique \cite{wolfnew2001,luthiphysical2005,hau24}. 
 Overtone polished 41${^\circ}$ X-cut LiNbO$_3$ resonance transducers are bonded with Thiokol LP-32 to the FIB-prepared parallel sample surfaces. We performed the ultrasound experiments at  a frequency of about 30 MHz. 
We chose the propagation direction $k \parallel b$ with polarization $u \parallel a$, corresponding to the elastic mode $(c_{11} - c_{12})/2$ in the hexagonal lattice [Fig. \ref{fig:CrystalStructure}(b)]. 
The sound velocity, $v$, is related to the concomitant elastic constant $c_{ij} = \rho v^2$, where $\rho$ is the mass density of the material. The sound velocity is obtained from the time delay of the acoustic signal in the sample.

\begin{figure}
	\centering
	\includegraphics[]{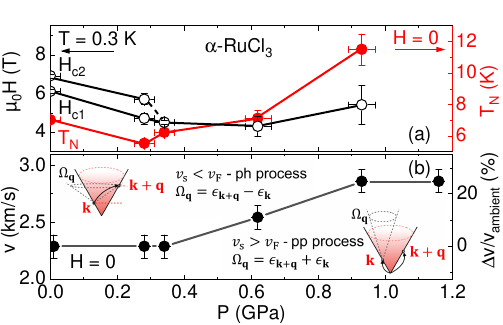}
	\caption{(a) Pressure dependence of the critical temperature $T_N$ (red solid circles, right scale) in zero magnetic field and critical magnetic fields $H_{c1}$ and $H_{c2}$ (black open circles, left scale) for $H \parallel a$ at 0.3 K in $\alpha$-RuCl$_3$.
	(b) Pressure dependence of the measured sound velocity $v$ of the $(c_{11}-c_{12})/2$ acoustic mode and relative change $[v(P)- v_{\mathrm{ambient}}]/v_{\mathrm{ambient}}$. Insets in (b) show two  different phonon scattering processes by Majorana fermions (see text for details). \label{fig:Tc_Hc_v_vs_p} }
\end{figure}

\subsection{Pressure cell}

We used a commercially available cylinder-piston cell from C\&T Factory, model CTF-HHPC40 with Daphne 7373 as the pressure medium \cite{staskopressure2020}, adapted for ultrasound experiments following the method described in Refs. \cite{mombetsustudy2016,kepapiston2016}. 
The pressure cell was thermally linked to the $^3$He pot of a commercial $^3$He cryostat via a copper rod.
We attached a calibrated RuO$_2$ temperature sensor to the pressure cell. The ultrasound experiments in applied magnetic field are performed under zero-field-cooled (ZFC) conditions.
For preparing the ultrasound measurement under pressure, we fixed the crystal on a sapphire platform, which is attached to the outer conductor of the coaxial cable in the pressure cell. 
Up to the highest pressure of 1.16 GPa, we observe multiple ultrasound echoes due to multiple propagation and signal reflections in the sample. Nevertheless, because of the irregular shape of the sample, we focus on the  $0^{\mathrm{th}}$ echo (a single sound transmission through the sample) to avoid a superposition with the delayed signals reflected from the sample sides. The pressure is defined from the pressure-dependent superconducting transition in tin \cite{eilingpressure1981}. We changed the pressure at room temperature. With slowly cooling the cell to cryogenic temperatures, the hydrostatic conditions remain valid down to 0.3 K, even though a crystallization of daphne 7373 occurs \cite{yokojap07}.

\begin{figure}
	\centering
	\includegraphics[]{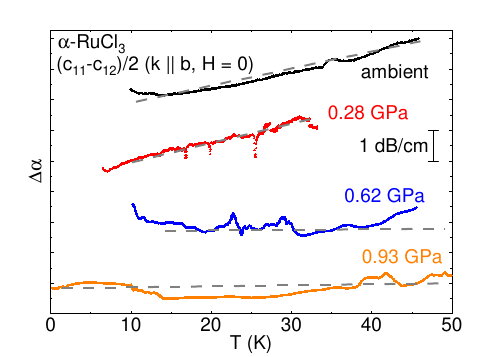}
	\caption{The attenuation of the transverse acoustic
		mode $(c_{11}-c_{12})/2$  ($k \parallel b$, $u \parallel a$) versus temperature in $\alpha$-RuCl$_3$ at some selected values of the hydrostatic pressure in zero magnetic field beyond the ordered magnetic state. The curves are arbitrarily shifted along the $y$ axis for clarity. The ambient-pressure curve is taken from Ref. \cite{hau24}. The gray dashed lines are guides to the eye. The low-temperature data were removed close to and below $T_N$ to mask the dominant attenuation anomaly at $T_N$ \cite{hau24}.
		\label{fig:Attenuation}}
\end{figure}

\begin{figure*}
	\centering
	\includegraphics[]{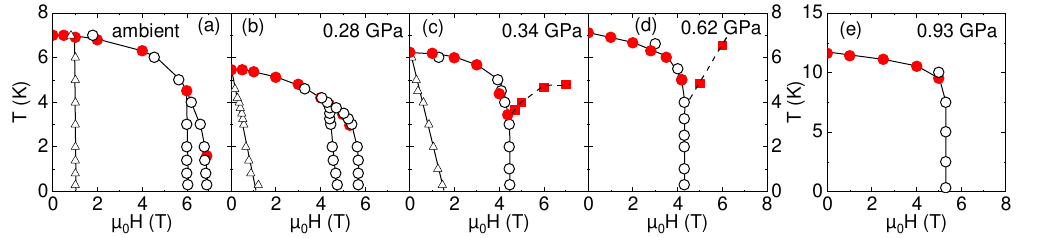}
	\caption{(a) - (e) $H$-$T$ phase diagrams ($H   \parallel a$)  of $\alpha$-RuCl$_3$ at various hydrostatic pressures. We extracted critical temperatures from temperature dependent measurements (red circles) and critical fields from field-dependent measurements (open circles and triangles) from our ultrasound results shown in Figs. \ref{fig:Tdep_Hdep}(a) and \ref{fig:Tdep_Hdep}(b). No transitions have been observed at 1.16 GPa. See text for details.
	\label{fig:PhaseDiagram}}
\end{figure*}

\section{Results and Discussion}
\subsection{Change of the sound velocity under hydrostatic pressure}

Figure \ref{fig:Tdep_Hdep} shows the temperature and magnetic-field dependence of the relative sound velocity change, $\Delta v/v$, for the transverse acoustic mode $(c_{11}-c_{12})/2$ measured at various hydrostatic pressures. At ambient pressure, this mode exhibits pronounced softening,
near the AFM ordering temperature, $T_{N}$ [Fig. \ref{fig:Tdep_Hdep}(a)]. Similar anomalies are observed at magnetic transitions in applied magnetic field, specifically at the critical fields $H_{c1}$ and $H_{c2}$ [Fig. \ref{fig:Tdep_Hdep}(b)] \cite{hau24}. Much smaller kink-like anomalies are detected in the sound velocity below 2 T, where a magnetic domain rearrangement occurs \cite{hau24,sea17}.

As pressure increases,  the anomalies in $\Delta v/v$
are significantly modified and their position changes.
Specifically, at 0.62 GPa, the sound velocity exhibits a pronounced increase both versus temperature and magnetic field, contrasting with the minima observed at lower pressures.  Additionally, the anomaly becomes broader, with the increase in relative velocity occurring over a wider temperature range. At this pressure, the sound velocity increases significantly
($\approx$ 4 \%) and is accompanied by a strong hysteresis in the field dependence below 4 T, suggesting the presence of a metastable magnetic state. In this state, the dimerized and the non-dimerized phases can coexist, consistent with the strongly first-order nature of the dimerization transition. At 0.93 GPa, only a small, broad step-like anomaly is observed in both the temperature and magnetic-field dependences of the sound velocity [Figs.~\ref{fig:Tdep_Hdep}(a) and \ref{fig:Tdep_Hdep}(b)].

Anomalies in the temperature and magnetic-field dependence of the sound velocity, signaling phase transitions, are indicated with arrows in Fig. \ref{fig:Tdep_Hdep}. Accordingly, $T_{N}$, $H_{c2}$, and $H_{c1}$ are plotted as a function of pressure in Fig. \ref{fig:Tc_Hc_v_vs_p}(a). First, $T_{N}$ shows a slight decrease with increasing pressure but rises at higher pressures, ultimately surpassing its ambient-pressure value. These trends in $T_{N}$ are consistent with results reported in Ref. \cite{cui17}.
Remarkably, the absolute value of the sound velocity increases significantly between 0.34 and 0.93 GPa [Fig. \ref{fig:Tc_Hc_v_vs_p}(b)] indicating a notable stiffening of the $(c_{11}-c_{12})/2$ elastic mode in this pressure range \footnote{We note that the sample size undergoes a change under applied pressure. At room temperature, we verified that the relative length change within the hexagonal plane remains below 3\% ($\Delta l/l_0 < 3\%$) for pressures up to 1 GPa. Based on XRD analysis \cite{bas18}, the sample-size change under pressure may be larger at lower temperatures. Since we do not have precise values, we did not include this correction in our analysis. We acknowledge that the reported sound velocity and its relative change might be slightly overestimated.}. 

\subsection{Ultrasound attenuation}
We now consider the effects of hydrostatic pressure on the sound attenuation in $\alpha$-RuCl$_3$. 
Figures \ref{fig:Tdep_Hdep}(c) and \ref{fig:Tdep_Hdep}(d) show the sound attenuation at selected pressure values versus temperature and magnetic field, respectively. The sound attenuation exhibits pronounced anomalies at the phase transitions. There is an attenuation maximum at $T_{N}$, $H_{c2}$, and $H_{c1}$ at low pressure values, which transforms in more sophisticated features with pressure increase. The position of the attenuation anomalies matches well with the sound-velocity anomalies. Consistent with the sound-velocity observations, the sound attenuation undergoes a significant change at 0.62 GPa.

 Figure \ref{fig:Attenuation} shows in enlarged scale the temperature dependence of the sound attenuation of $(c_{11}-c_{12})/2$ at pressure values 0, 0.28, 0.62, and 0.93 GPa in zero magnetic field. 
To highlight the subtle linear temperature dependence discussed later, we have masked, for clarity, the data points corresponding to the large anomalies associated with the transition into the magnetically ordered state [Figs. \ref{fig:Tdep_Hdep}(c) and \ref{fig:Tdep_Hdep}(d)].

At ambient pressure, the attenuation exhibits a linear temperature dependence with a non-zero slope. This is consistent with the notion that the sound attenuation has contributions from microscopic processes, where a phonon scatters a positive-energy fermionic excitation of the nearby QSL to a  higher-energy fermion state, dubbed particle-hole channel \cite{yephonon2020, metavitsiadisphonon2020, hau24}.  This scattering mechanism is dominant when the sound velocity is smaller than the characteristic Fermi velocity $v_F$ of the fractionalized excitations (discussed further in Sec. \ref{Discussion}). As the pressure increases, the slope of the linear temperature dependence decreases. At 0.62 GPa, and discarding noise in the data, the sound attenuation becomes nearly temperature independent. A similar, almost flat temperature profile of the sound attenuation is observed at 0.93 GPa.

In Sec. \ref{Discussion}, as a main point of this work, we rationalize this change of slope of the attenuation to arise from a crossover of the scattering of phonons from particle-hole- to particle-particle-like processes in the proximate QSL \cite{yephonon2020, metavitsiadisphonon2020, hau24}. On the one hand, this is consistent with the experimentally observed increase of the sound velocity [see  Fig. \ref{fig:Tc_Hc_v_vs_p}(b)]. On the other hand, pressure also modifies the effective magnetic exchange couplings. This can further promote a change of the scattering mechanism. Both, first-principles calculations \cite{Kaib2021,wolf22} and quantum-chemistry analyses \cite{Yadav2018} have highlighted a significant dependence of magnetic coupling constants on pressure. The various magnetic interactions in $\alpha$-RuCl$_3$ respond very differently to lattice variations, with the Kitaev interaction being the most sensitive. We examine these trends in detail in Sec. \ref{Discussion}. We will show, that not only the variation of the sound velocity, but also of the Majorana-fermion velocity is a key parameter for the sound attenuation in $\alpha$-RuCl$_3$ under pressure.

\begin{figure*}
	\centering
\includegraphics[width=1.0\textwidth]{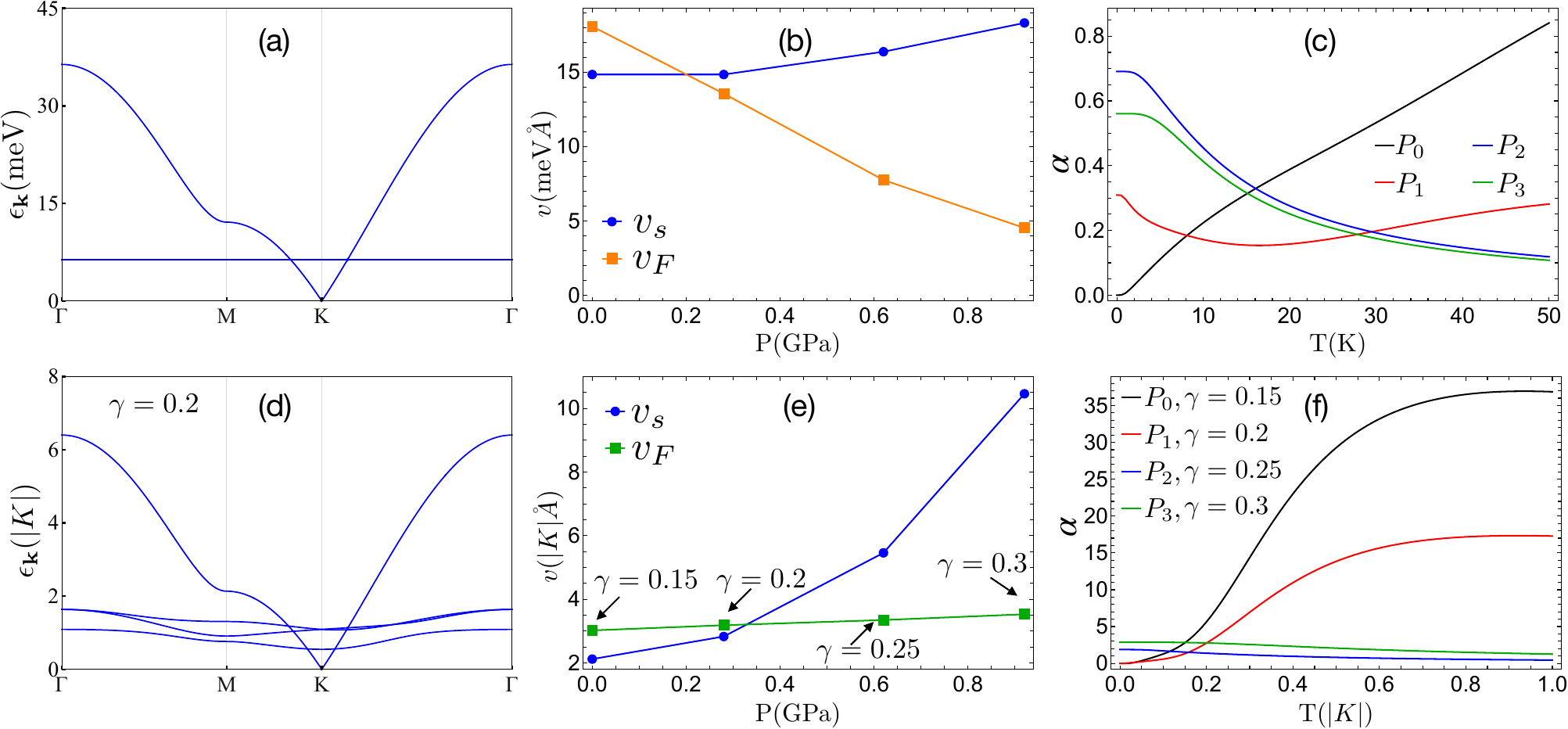}
\caption{Spectrum of fermionic excitations for (a) the Kitaev model with $K = -7$ meV  and (d) the $J$-$K$-$\Gamma$ model with $\gamma = -J/|K| = \Gamma/|K| = 0.2$. Panels (b) and (e) compare the experimentally measured sound velocity, $v_s$, with the theoretically estimated Fermi velocity, $v_F$, for the two models as a function of pressure. 
Panels (c) and (f) present the computed sound attenuation of the transverse acoustic mode $(c_{11}-c_{12})/2$ versus temperature for various pressures for the pure Kitaev model and the $J$-$K$-$\Gamma$ model, respectively.
 In (c), the attenuation is due to only ph-processes at $P_0$,  and only due to pp-processes at $P_1$, $P_2$, and $P_3$. In (f), the attenuation of phonons is dominated by ph-processes at $P_0$ and $P_1$ and by pp-processes at $P_2$ and $P_3$. 
Our calculations follow the pressure dependence of exchange couplings as suggested by ab-initio calculations \cite{wolf22}. In (b) and (c), the pure Kitaev model incorporates the pressure-induced reduction of the Kitaev interaction, scaled to $K = -7$ meV at ambient pressure. In (e) and (f), the $J$-$K$-$\Gamma$ model adopts the pressure-dependent ratio $\gamma = -J/|K| = \Gamma/|K|$ from \cite{wolf22}, while ensuring the system remains in the quantum-spin-liquid state (see text for details).
 In (d)-(f), all quantities (energies, temperatures, velocities) are expressed in units of $|K|$, whereas (a)-(c) use physical units.
  \label{fig:theory}}
\end{figure*}

\subsection{
Pressure dependent $H$-$T$ phase diagram}

Using the positions of the acoustic anomalies related to the phase transitions (Fig. \ref{fig:Tdep_Hdep}), we construct the $H$-$T$ phase diagrams of $\alpha$-RuCl$_3$ for selected pressures, as shown in Fig. \ref{fig:PhaseDiagram}. 
 
The $H$-$T$ phase diagram at ambient pressure shown in Fig.~\ref{fig:PhaseDiagram}(a) agrees with the $H$-$T$ phase diagram we reported in Ref. \cite{hau24}, except that the crossover regime beyond the ordered state related to the spin-gap above $H_{c2}$, detected by the acoustic mode $c_{11}$ \cite{hau24}, is not observed in $(c_{11}-c_{12})/2$ reported here.
As pressure increases, the $H$-$T$ phase diagram undergoes significant modifications.
  The low-field anomaly near 1 T, related to domain rearrangements in magnetic field \cite{sea17}, is suppressed at 0.62 GPa.
 Additionally, the transition at $H_{c1}$ indicating a change in the magnetic structure between zigzag phases with different stacking orders \cite{balzfield2021,schoenemannthermal2020,bal19} is not detected above 0.28 GPa, suggesting that $H_{c1}$ and $H_{c2}$ may merge around 0.3 GPa. 
 The anomalies marked by red squares at 0.34 and 0.62 GPa might be related to remnants of phases with different stacking orders, 
although this interpretation remains uncertain. Above 0.93 GPa, no evidence of 3D magnetic order is detected.

These findings highlight that pressure significantly alters the magnetic landscape, suppressing specific transitions, merging critical fields, and eliminating 3D magnetic order. In turn, the response to pressure reveals an intricate interplay between stacking order, spin gaps, and magnetic transitions.

\section{Discussion}\label{Discussion}

As we discussed previously \cite{hau24}, the temperature and in-plane magnetic-field dependence of the sound attenuation in $\alpha$-RuCl$_3$ is consistent 
with  theoretical predictions that phonons are scattered by fractionalized excitations arising from the proximity of $\alpha$-RuCl$_3$ to a Kitaev  QSL  \cite{yephonon2020, metavitsiadisphonon2020, fengtemperature2021,sin23,Singh2024, Dantas2024}.
 We showed \cite{hau24}, that the sound attenuation is  linear in temperature when
the  phonon velocity $v_s$ is smaller than  the characteristic velocity of the low-energy fermionic excitations, $v_F$, as observed for the transverse phonon mode,  $(c_{11}-c_{12})/2$. Moreover, the attenuation is almost temperature independent with a slight downturn in the case of longitudinal  mode $c_{11}$, which is consistent with a sound velocity slightly exceeding $v_F$. These different temperature dependences of the sound attenuation of the longitudinal and transverse phonon mode arise from the underlying scattering mechanisms. For $v_s<v_F$, a phonon excites an occupied fermion state to a higher-energy state [particle-hole (ph) channel]. In contrast, for $v_s>v_F$, the dominant process involves the phonon decaying into two fermions, both with positive energy [particle-particle (pp) channel] \cite{yephonon2020, metavitsiadisphonon2020}.

Here, we start by investigating the temperature dependence of the sound attenuation in the pure Kitaev model, incorporating a pressure-dependent variation of the Kitaev interaction as proposed in Ref. \cite{wolf22}. This study demonstrates that the Kitaev interaction is highly sensitive to applied pressure. 
 At ambient pressure, it is the dominant interaction in the system and it is ferromagnetic in nature. However, as pressure increases, the Kitaev interaction acquires a strong positive contribution, 
  eventually leading to a sign change at sufficiently high pressures. Within the pressure range considered in this study (up to approximately 1 GPa), the dominant effect of pressure is a reduction in the magnitude of the Kitaev interaction.

Our results for   the pure Kitaev model are presented in Figs. \ref{fig:theory}(a) - \ref{fig:theory}(c). In panel (a),  the spectrum of fermionic excitations for the Kitaev model is shown, computed with 
 $K = -7$ meV  at ambient pressure.
 The characteristic dispersing modes, arising from free Majoranas hopping on the lattice,  feature the Dirac cones at the $K$ points of the Brillouin zone. 
 Additionally, the flat bands correspond to  the static $Z_2$ fluxes. 
 Panels (b) and (c) present the computed Fermi velocity and sound attenuation, respectively, for varying values of the Kitaev interaction. Based on the trend reported in Ref. \cite{wolf22}, the Kitaev interaction values $K$ are taken to be -7, -5.25, -3, and  -1.75 meV for the four pressure values $P_0=0$, $P_1=0.28$ GPa, $P_2=0.62$ GPa, and $P_3=0.93$, respectively.

In Fig. \ref{fig:theory}(b), the Fermi velocity $v_F$ is compared with the experimental sound velocity shown in Fig. \ref{fig:Tc_Hc_v_vs_p}(b). We estimate $v_F$ from the slope
of the Dirac cones. At  ambient pressure, $v_F \simeq$ 2746 m/s 
(18 meV \AA),  consistent with previous estimates \cite{lidivergence2021}. 
With increasing pressure, $v_F$ decreases and, just above 0.2 GPa, becomes smaller than the sound velocity. 
This marks a change in the phonon-attenuation mechanism, shifting from ph-scattering to pp-scattering. This change is evident in the sound-attenuation behavior shown in Fig. \ref{fig:theory}(c).
 We see a monotonically increasing attenuation at $P_0$, which arises from pure ph-scattering of the acoustic phonons in the regime $v_s<v_F$. At $P_1$, $v_s$ is already  slightly larger than $v_F$, 
 placing the system in the pp-scattering regime, where attenuation decreases with temperature as low-energy states become occupied. However, we observe at $P_1$ = 0.28 GPa an increase in attenuation for temperatures above 10 K (Fig. \ref{fig:Attenuation}).
 This arises due to ph-scattering, which becomes significant as higher-energy states start to populate. Although we estimate $v_s>v_F$,
 the difference between the two velocities is very small, allowing a ph-contribution from higher-energy states away from the fermionic cone.
 At $P_2$ and $P_3$, the system is well within the pp-scattering regime. 

The experimentally observed change in the temperature dependence of the sound attenuation occurs at slightly higher pressures (see 
Fig. \ref{fig:Attenuation}). Using a more suitable ansatz, we compute the sound attenuation within the $J$-$K$-$\Gamma$ model [Figs. \ref{fig:theory}(d)-\ref{fig:theory}(f)], 
in which we incorporated  
the pressure-dependent variation of the coupling constants.
Since the  generic $J$-$K$-$\Gamma$ model is not exactly solvable, our calculations are performed using a self-consistent slave-fermion mean-field approach, as described in Ref.~\cite{sin23}. 
 The spectrum of the $J$-$K$-$\Gamma$ model  with $\gamma$ = $-J/|K|$ = $\Gamma/|K| = 0.2$ is shown in  Fig. \ref{fig:theory}(d). 
The original dispersing mode is only slightly modified, with the Dirac cones remaining at the $K$ points  and the Fermi velocity of the Dirac cones slightly decreasing. However, the flux bands loose their degeneracy and develop a small, yet noticeable dispersion.
 
 We note that while extending the analysis beyond the pure Kitaev model, we still perform our calculations for parameters of the 
  $J$-$K$-$\Gamma$ model, which correspond  to the spin-liquid regime. 
 Specifically, we  set $\gamma=0.2$, which provides a more realistic representation of the coupling parameters in $\alpha$-RuCl$_3$ under low-pressure conditions compared to the pure Kitaev model. Nonetheless, we emphasize that this choice does not yield a fully quantitative description of $\alpha$-RuCl$_3$.
 
In Figs. \ref{fig:theory}(e) and \ref{fig:theory}(f), we use the 
overall  trend  for the change of the  exchange parameters reported in \cite{wolf22}. According to this trend, the magnitude of the Kitaev interaction $K$ decreases with increasing pressure, while the Heisenberg interaction $J$ and the off-diagonal interaction $\Gamma$ are enhanced. In Fig. \ref{fig:theory}(e), the sound velocity $v_s$ and the Fermi velocity $v_F$ are plotted for different values of $\gamma$. We assume that $\gamma$, 
representing the relative strengths of the Heisenberg interaction $J$ and off-diagonal interaction $\Gamma$, increases with pressure from  $P_0$  to $P_3$.  In this case, the sound velocity $v_s$ surpasses the Fermi velocity
$v_F$ at a pressure slightly higher, as estimated in the pure Kitaev model.
   
The resulting sound attenuation is presented in Fig. \ref{fig:theory}(f). Similar to the pure Kitaev model, the sound attenuation at the lowest two pressures, $P_0$ and $P_1$, exhibits a roughly linear temperature dependence over a certain range of temperatures, followed by saturation and a subsequent decrease at high temperatures (not shown). At higher pressures, $P_2$ and $P_3$, the sound attenuation becomes nearly temperature independent. This behavior can be attributed to the dominant scattering mechanisms: At $P_0$ and $P_1$, the attenuation is primarily governed by ph-processes, whereas at $P_2$ and $P_3$, it is dominated entirely by pp-scattering.

Overall, we obtain that in both, the pure Kitaev and the $J\text{-}K\text{-}\Gamma$ models, the sound attenuation changes from a (quasi-) linear increase with temperature to a (weakly) decreasing or almost temperature-independent behavior as pressure is increased. This is similar as observed in experiment, as shown in Fig.~\ref{fig:Attenuation}. We strongly emphasize, that our aim is not to quantitatively describe the variation with pressure of the measured sound attenuation. Instead, and far more important, independent of the model used, we find a pressure-induced crossing, above which the calculated Fermi velocity of the QSL quasiparticles becomes smaller as the experimentally determined sound velocity. This can favorably explain the observed temperature dependence of the sound attenuation (Fig. \ref{fig:Attenuation}). Fine details of the temperature and pressure variations of $\Delta\alpha$ are certainly beyond our calculations.

\section{Summary}

We have presented an ultrasound study of $\alpha$-RuCl$_3$ under hydrostatic pressures up to 1.16 GPa. Anomalies in the sound velocity allows us to extract the $H$-$T$ phase diagrams over a wide range of pressures. At ambient pressure, we identify a marked softening of the acoustic mode $(c_{11}-c_{12})/2$ at $T_N$, highlighting strong magnetoelastic coupling between lattice vibrations and magnetic degrees of freedom. We find this coupling to persists at temperatures well above the transition into the disordered state  [Fig. \ref{fig:Tdep_Hdep}(a)].

The linear temperature dependence of the sound attenuation of $(c_{11}-c_{12})/2$ at ambient pressure beyond the ordered state is consistent with phonons scattered by Majorana fermions in a process dominated by particle-hole excitations. As pressures increases, however, the sound attenuation becomes nearly temperature independent, suggesting a crossover of the scattering channel from a ph-dominated to rather a pp-regime. This crossover can be understood by a reversal of the magnitudes of the sound and Fermi velocities from $v_s < v_F$ at low pressure to $v_s > v_F$ above 0.28 GPa. This reversal is of twofold origin. I.e., from our experimental findings, the measured sound velocity of $(c_{11}-c_{12})/2$ increases substantially with pressure, while from our theoretical analysis, the Fermi velocities of an effective $J$-$K$-$\Gamma$ model remain almost invariant with pressure.

Overall, our findings provide compelling evidence for the tunability of phonon dynamics in $\alpha$-RuCl$_3$ through hydrostatic pressure and 
further improve our understanding of spin-strain interactions in Kitaev magnets.

\begin{acknowledgments}
 We gratefully acknowledge our previous collaboration with  Mengxing Ye and Peter Stavropoulos and discussions with Roser Valenti, which significantly contributed to shaping our understanding and approach in this work.
We acknowledge support of the HLD at HZDR, member of the European Magnetic Field Laboratory (EMFL). Work of S.Z., A.H., W.B., and J.W. has been supported in part by the DFG through SFB 1143 (project-id 247310070). Work of A.H., J.W., and S.Z. has been supported in part by the DFG trough excellence cluster $ct.qmat$ (EXC 2147, project-id 39085490). W.B. acknowledges kind hospitality of the PSM, Dresden. The work of N.B.P. and S.S. was supported by the U.S. Department of Energy, Office of Science, Basic Energy Sciences under Award No. DE-SC0018056.  N.B.P.  also acknowledges the hospitality and partial support  of the Technical University of Munich – Institute for Advanced Study.
The work of V.T. was supported by the DFG through Transregional Research Collaboration TRR 80 (Augsburg, Munich, and Stuttgart) as well as by the project ANCD 20.80009.5007.19 (Moldova).
\end{acknowledgments}

\bibliography{2023_RuCl3_Pressure_bibliography.bib}

\begin{thebibliography}{76}%
\makeatletter
\providecommand \@ifxundefined [1]{%
 \@ifx{#1\undefined}
}%
\providecommand \@ifnum [1]{%
 \ifnum #1\expandafter \@firstoftwo
 \else \expandafter \@secondoftwo
 \fi
}%
\providecommand \@ifx [1]{%
 \ifx #1\expandafter \@firstoftwo
 \else \expandafter \@secondoftwo
 \fi
}%
\providecommand \natexlab [1]{#1}%
\providecommand \enquote  [1]{``#1''}%
\providecommand \bibnamefont  [1]{#1}%
\providecommand \bibfnamefont [1]{#1}%
\providecommand \citenamefont [1]{#1}%
\providecommand \href@noop [0]{\@secondoftwo}%
\providecommand \href [0]{\begingroup \@sanitize@url \@href}%
\providecommand \@href[1]{\@@startlink{#1}\@@href}%
\providecommand \@@href[1]{\endgroup#1\@@endlink}%
\providecommand \@sanitize@url [0]{\catcode `\\12\catcode `\$12\catcode
  `\&12\catcode `\#12\catcode `\^12\catcode `\_12\catcode `\%12\relax}%
\providecommand \@@startlink[1]{}%
\providecommand \@@endlink[0]{}%
\providecommand \url  [0]{\begingroup\@sanitize@url \@url }%
\providecommand \@url [1]{\endgroup\@href {#1}{\urlprefix }}%
\providecommand \urlprefix  [0]{URL }%
\providecommand \Eprint [0]{\href }%
\providecommand \doibase [0]{https://doi.org/}%
\providecommand \selectlanguage [0]{\@gobble}%
\providecommand \bibinfo  [0]{\@secondoftwo}%
\providecommand \bibfield  [0]{\@secondoftwo}%
\providecommand \translation [1]{[#1]}%
\providecommand \BibitemOpen [0]{}%
\providecommand \bibitemStop [0]{}%
\providecommand \bibitemNoStop [0]{.\EOS\space}%
\providecommand \EOS [0]{\spacefactor3000\relax}%
\providecommand \BibitemShut  [1]{\csname bibitem#1\endcsname}%
\let\auto@bib@innerbib\@empty
\bibitem [{\citenamefont {Jackeli}\ and\ \citenamefont
  {Khaliullin}(2009)}]{jackeli09}%
  \BibitemOpen
  \bibfield  {author} {\bibinfo {author} {\bibfnamefont {G.}~\bibnamefont
  {Jackeli}}\ and\ \bibinfo {author} {\bibfnamefont {G.}~\bibnamefont
  {Khaliullin}},\ }\bibfield  {title} {\bibinfo {title} {{Mott Insulators in
  the Strong Spin-Orbit Coupling Limit: From Heisenberg to a Quantum Compass
  and Kitaev Models}},\ }\href {https://doi.org/10.1103/PhysRevLett.102.017205}
  {\bibfield  {journal} {\bibinfo  {journal} {Phys. Rev. Lett.}\ }\textbf
  {\bibinfo {volume} {102}},\ \bibinfo {pages} {017205} (\bibinfo {year}
  {2009})}\BibitemShut {NoStop}%
\bibitem [{\citenamefont {Takagi}\ \emph {et~al.}(2019)\citenamefont {Takagi},
  \citenamefont {Takayama}, \citenamefont {Jackeli}, \citenamefont
  {Khaliullin},\ and\ \citenamefont {Nagler}}]{takagiconcept2019}%
  \BibitemOpen
  \bibfield  {author} {\bibinfo {author} {\bibfnamefont {H.}~\bibnamefont
  {Takagi}}, \bibinfo {author} {\bibfnamefont {T.}~\bibnamefont {Takayama}},
  \bibinfo {author} {\bibfnamefont {G.}~\bibnamefont {Jackeli}}, \bibinfo
  {author} {\bibfnamefont {G.}~\bibnamefont {Khaliullin}},\ and\ \bibinfo
  {author} {\bibfnamefont {S.~E.}\ \bibnamefont {Nagler}},\ }\bibfield  {title}
  {\bibinfo {title} {Concept and realization of {Kitaev} quantum spin
  liquids},\ }\href {https://doi.org/10.1038/s42254-019-0038-2} {\bibfield
  {journal} {\bibinfo  {journal} {Nat. Rev. Phys.}\ }\textbf {\bibinfo {volume}
  {1}},\ \bibinfo {pages} {264} (\bibinfo {year} {2019})}\BibitemShut {NoStop}%
\bibitem [{\citenamefont {Trebst}\ and\ \citenamefont
  {Hickey}(2022)}]{Trebst2022}%
  \BibitemOpen
  \bibfield  {author} {\bibinfo {author} {\bibfnamefont {S.}~\bibnamefont
  {Trebst}}\ and\ \bibinfo {author} {\bibfnamefont {C.}~\bibnamefont
  {Hickey}},\ }\bibfield  {title} {\bibinfo {title} {Kitaev materials},\ }\href
  {https://doi.org/https://doi.org/10.1016/j.physrep.2021.11.003} {\bibfield
  {journal} {\bibinfo  {journal} {Phys. Rep.}\ }\textbf {\bibinfo {volume}
  {950}},\ \bibinfo {pages} {1} (\bibinfo {year} {2022})}\BibitemShut {NoStop}%
\bibitem [{\citenamefont {Rousochatzakis}\ \emph {et~al.}(2024)\citenamefont
  {Rousochatzakis}, \citenamefont {Perkins}, \citenamefont {Luo},\ and\
  \citenamefont {Kee}}]{Rousochatzakis2024}%
  \BibitemOpen
  \bibfield  {author} {\bibinfo {author} {\bibfnamefont {I.}~\bibnamefont
  {Rousochatzakis}}, \bibinfo {author} {\bibfnamefont {N.~B.}\ \bibnamefont
  {Perkins}}, \bibinfo {author} {\bibfnamefont {Q.}~\bibnamefont {Luo}},\ and\
  \bibinfo {author} {\bibfnamefont {H.-Y.}\ \bibnamefont {Kee}},\ }\bibfield
  {title} {\bibinfo {title} {{Beyond Kitaev physics in strong spin-orbit
  coupled magnets}},\ }\href {https://doi.org/10.1088/1361-6633/ad208d}
  {\bibfield  {journal} {\bibinfo  {journal} {Reports on Progress in Physics}\
  }\textbf {\bibinfo {volume} {87}},\ \bibinfo {pages} {026502} (\bibinfo
  {year} {2024})}\BibitemShut {NoStop}%
\bibitem [{\citenamefont {Kitaev}(2006)}]{kitaevanyons2006}%
  \BibitemOpen
  \bibfield  {author} {\bibinfo {author} {\bibfnamefont {A.}~\bibnamefont
  {Kitaev}},\ }\bibfield  {title} {\bibinfo {title} {Anyons in an exactly
  solved model and beyond},\ }\href
  {https://doi.org/https://doi.org/10.1016/j.aop.2005.10.005} {\bibfield
  {journal} {\bibinfo  {journal} {Annals of Physics}\ }\textbf {\bibinfo
  {volume} {321}},\ \bibinfo {pages} {2} (\bibinfo {year} {2006})},\ \bibinfo
  {note} {january Special Issue}\BibitemShut {NoStop}%
\bibitem [{\citenamefont {Plumb}\ \emph {et~al.}(2014)\citenamefont {Plumb},
  \citenamefont {Clancy}, \citenamefont {Sandilands}, \citenamefont {Shankar},
  \citenamefont {Hu}, \citenamefont {Burch}, \citenamefont {Kee},\ and\
  \citenamefont {Kim}}]{Plumb2014}%
  \BibitemOpen
  \bibfield  {author} {\bibinfo {author} {\bibfnamefont {K.~W.}\ \bibnamefont
  {Plumb}}, \bibinfo {author} {\bibfnamefont {J.~P.}\ \bibnamefont {Clancy}},
  \bibinfo {author} {\bibfnamefont {L.~J.}\ \bibnamefont {Sandilands}},
  \bibinfo {author} {\bibfnamefont {V.~V.}\ \bibnamefont {Shankar}}, \bibinfo
  {author} {\bibfnamefont {Y.~F.}\ \bibnamefont {Hu}}, \bibinfo {author}
  {\bibfnamefont {K.~S.}\ \bibnamefont {Burch}}, \bibinfo {author}
  {\bibfnamefont {H.-Y.}\ \bibnamefont {Kee}},\ and\ \bibinfo {author}
  {\bibfnamefont {Y.-J.}\ \bibnamefont {Kim}},\ }\bibfield  {title} {\bibinfo
  {title} {\textit{{$\ensuremath{\alpha}-{\mathrm{RuCl}}_{3}$: {A}
  {Spin}-{Orbit} {Assisted} {Mott} {Insulator} on a {Honeycomb} {Lattice}}}},\
  }\href {https://doi.org/10.1103/PhysRevB.90.041112} {\bibfield  {journal}
  {\bibinfo  {journal} {Phys. Rev. B}\ }\textbf {\bibinfo {volume} {90}},\
  \bibinfo {pages} {041112(R)} (\bibinfo {year} {2014})}\BibitemShut {NoStop}%
\bibitem [{\citenamefont {Sears}\ \emph {et~al.}(2015)\citenamefont {Sears},
  \citenamefont {Songvilay}, \citenamefont {Plumb}, \citenamefont {Clancy},
  \citenamefont {Qiu}, \citenamefont {Zhao}, \citenamefont {Parshall},\ and\
  \citenamefont {Kim}}]{Sears2015}%
  \BibitemOpen
  \bibfield  {author} {\bibinfo {author} {\bibfnamefont {J.~A.}\ \bibnamefont
  {Sears}}, \bibinfo {author} {\bibfnamefont {M.}~\bibnamefont {Songvilay}},
  \bibinfo {author} {\bibfnamefont {K.~W.}\ \bibnamefont {Plumb}}, \bibinfo
  {author} {\bibfnamefont {J.~P.}\ \bibnamefont {Clancy}}, \bibinfo {author}
  {\bibfnamefont {Y.}~\bibnamefont {Qiu}}, \bibinfo {author} {\bibfnamefont
  {Y.}~\bibnamefont {Zhao}}, \bibinfo {author} {\bibfnamefont {D.}~\bibnamefont
  {Parshall}},\ and\ \bibinfo {author} {\bibfnamefont {Y.-J.}\ \bibnamefont
  {Kim}},\ }\bibfield  {title} {\bibinfo {title} {\textit{{Magnetic Order in
  $\ensuremath{\alpha}-{\text{RuCl}}_{3}$: A Honeycomb-Lattice Quantum Magnet
  with Strong Spin-Orbit Coupling}}},\ }\href
  {https://doi.org/10.1103/PhysRevB.91.144420} {\bibfield  {journal} {\bibinfo
  {journal} {Phys. Rev. B}\ }\textbf {\bibinfo {volume} {91}},\ \bibinfo
  {pages} {144420} (\bibinfo {year} {2015})}\BibitemShut {NoStop}%
\bibitem [{\citenamefont {Kasahara}\ \emph {et~al.}(2018)\citenamefont
  {Kasahara}, \citenamefont {Ohnishi}, \citenamefont {Mizukami}, \citenamefont
  {Tanaka}, \citenamefont {Ma}, \citenamefont {Sugii}, \citenamefont {Kurita},
  \citenamefont {Tanaka}, \citenamefont {Nasu}, \citenamefont {Motome},
  \citenamefont {Shibauchi},\ and\ \citenamefont
  {Matsuda}}]{kasaharamajorana2018}%
  \BibitemOpen
  \bibfield  {author} {\bibinfo {author} {\bibfnamefont {Y.}~\bibnamefont
  {Kasahara}}, \bibinfo {author} {\bibfnamefont {T.}~\bibnamefont {Ohnishi}},
  \bibinfo {author} {\bibfnamefont {Y.}~\bibnamefont {Mizukami}}, \bibinfo
  {author} {\bibfnamefont {O.}~\bibnamefont {Tanaka}}, \bibinfo {author}
  {\bibfnamefont {S.}~\bibnamefont {Ma}}, \bibinfo {author} {\bibfnamefont
  {K.}~\bibnamefont {Sugii}}, \bibinfo {author} {\bibfnamefont
  {N.}~\bibnamefont {Kurita}}, \bibinfo {author} {\bibfnamefont
  {H.}~\bibnamefont {Tanaka}}, \bibinfo {author} {\bibfnamefont
  {J.}~\bibnamefont {Nasu}}, \bibinfo {author} {\bibfnamefont {Y.}~\bibnamefont
  {Motome}}, \bibinfo {author} {\bibfnamefont {T.}~\bibnamefont {Shibauchi}},\
  and\ \bibinfo {author} {\bibfnamefont {Y.}~\bibnamefont {Matsuda}},\
  }\bibfield  {title} {\bibinfo {title} {Majorana quantization and half-integer
  thermal quantum {Hall} effect in a {Kitaev} spin liquid},\ }\href
  {https://doi.org/10.1038/s41586-018-0274-0} {\bibfield  {journal} {\bibinfo
  {journal} {Nature (London)}\ }\textbf {\bibinfo {volume} {559}},\ \bibinfo
  {pages} {227} (\bibinfo {year} {2018})}\BibitemShut {NoStop}%
\bibitem [{\citenamefont {Tanaka}\ \emph {et~al.}(2022)\citenamefont {Tanaka},
  \citenamefont {Mizukami}, \citenamefont {Harasawa}, \citenamefont
  {Hashimoto}, \citenamefont {Hwang}, \citenamefont {Kurita}, \citenamefont
  {Tanaka}, \citenamefont {Fujimoto}, \citenamefont {Matsuda}, \citenamefont
  {Moon},\ and\ \citenamefont {Shibauchi}}]{tanakathermodynamic2022}%
  \BibitemOpen
  \bibfield  {author} {\bibinfo {author} {\bibfnamefont {O.}~\bibnamefont
  {Tanaka}}, \bibinfo {author} {\bibfnamefont {Y.}~\bibnamefont {Mizukami}},
  \bibinfo {author} {\bibfnamefont {R.}~\bibnamefont {Harasawa}}, \bibinfo
  {author} {\bibfnamefont {K.}~\bibnamefont {Hashimoto}}, \bibinfo {author}
  {\bibfnamefont {K.}~\bibnamefont {Hwang}}, \bibinfo {author} {\bibfnamefont
  {N.}~\bibnamefont {Kurita}}, \bibinfo {author} {\bibfnamefont
  {H.}~\bibnamefont {Tanaka}}, \bibinfo {author} {\bibfnamefont
  {S.}~\bibnamefont {Fujimoto}}, \bibinfo {author} {\bibfnamefont
  {Y.}~\bibnamefont {Matsuda}}, \bibinfo {author} {\bibfnamefont {E.-G.}\
  \bibnamefont {Moon}},\ and\ \bibinfo {author} {\bibfnamefont
  {T.}~\bibnamefont {Shibauchi}},\ }\bibfield  {title} {\bibinfo {title}
  {Thermodynamic evidence for a field-angle-dependent {Majorana} gap in a
  {Kitaev} spin liquid},\ }\href {https://doi.org/10.1038/s41567-021-01488-6}
  {\bibfield  {journal} {\bibinfo  {journal} {Nat. Phys.}\ }\textbf {\bibinfo
  {volume} {18}},\ \bibinfo {pages} {429} (\bibinfo {year} {2022})}\BibitemShut
  {NoStop}%
\bibitem [{\citenamefont {Yokoi}\ \emph {et~al.}(2021)\citenamefont {Yokoi},
  \citenamefont {Ma}, \citenamefont {Kasahara}, \citenamefont {Kasahara},
  \citenamefont {Shibauchi}, \citenamefont {Kurita}, \citenamefont {Tanaka},
  \citenamefont {Nasu}, \citenamefont {Motome}, \citenamefont {Hickey},
  \citenamefont {Trebst},\ and\ \citenamefont
  {Matsuda}}]{yokoihalfinteger2021}%
  \BibitemOpen
  \bibfield  {author} {\bibinfo {author} {\bibfnamefont {T.}~\bibnamefont
  {Yokoi}}, \bibinfo {author} {\bibfnamefont {S.}~\bibnamefont {Ma}}, \bibinfo
  {author} {\bibfnamefont {Y.}~\bibnamefont {Kasahara}}, \bibinfo {author}
  {\bibfnamefont {S.}~\bibnamefont {Kasahara}}, \bibinfo {author}
  {\bibfnamefont {T.}~\bibnamefont {Shibauchi}}, \bibinfo {author}
  {\bibfnamefont {N.}~\bibnamefont {Kurita}}, \bibinfo {author} {\bibfnamefont
  {H.}~\bibnamefont {Tanaka}}, \bibinfo {author} {\bibfnamefont
  {J.}~\bibnamefont {Nasu}}, \bibinfo {author} {\bibfnamefont {Y.}~\bibnamefont
  {Motome}}, \bibinfo {author} {\bibfnamefont {C.}~\bibnamefont {Hickey}},
  \bibinfo {author} {\bibfnamefont {S.}~\bibnamefont {Trebst}},\ and\ \bibinfo
  {author} {\bibfnamefont {Y.}~\bibnamefont {Matsuda}},\ }\bibfield  {title}
  {\bibinfo {title} {Half-integer quantized anomalous thermal {Hall} effect in
  the {Kitaev} material candidate $\alpha$-{RuCl} $_{\textrm{3}}$},\ }\href
  {https://doi.org/10.1126/science.aay5551} {\bibfield  {journal} {\bibinfo
  {journal} {Science}\ }\textbf {\bibinfo {volume} {373}},\ \bibinfo {pages}
  {568} (\bibinfo {year} {2021})}\BibitemShut {NoStop}%
\bibitem [{\citenamefont {Balz}\ \emph {et~al.}(2021)\citenamefont {Balz},
  \citenamefont {Janssen}, \citenamefont {Lampen-Kelley}, \citenamefont
  {Banerjee}, \citenamefont {Liu}, \citenamefont {Yan}, \citenamefont
  {Mandrus}, \citenamefont {Vojta},\ and\ \citenamefont
  {Nagler}}]{balzfield2021}%
  \BibitemOpen
  \bibfield  {author} {\bibinfo {author} {\bibfnamefont {C.}~\bibnamefont
  {Balz}}, \bibinfo {author} {\bibfnamefont {L.}~\bibnamefont {Janssen}},
  \bibinfo {author} {\bibfnamefont {P.}~\bibnamefont {Lampen-Kelley}}, \bibinfo
  {author} {\bibfnamefont {A.}~\bibnamefont {Banerjee}}, \bibinfo {author}
  {\bibfnamefont {Y.~H.}\ \bibnamefont {Liu}}, \bibinfo {author} {\bibfnamefont
  {J.-Q.}\ \bibnamefont {Yan}}, \bibinfo {author} {\bibfnamefont {D.~G.}\
  \bibnamefont {Mandrus}}, \bibinfo {author} {\bibfnamefont {M.}~\bibnamefont
  {Vojta}},\ and\ \bibinfo {author} {\bibfnamefont {S.~E.}\ \bibnamefont
  {Nagler}},\ }\bibfield  {title} {\bibinfo {title} {Field-induced intermediate
  ordered phase and anisotropic interlayer interactions in
  $\ensuremath{\alpha}\text{\ensuremath{-}}{\mathrm{rucl}}_{3}$},\ }\href
  {https://doi.org/10.1103/PhysRevB.103.174417} {\bibfield  {journal} {\bibinfo
   {journal} {Phys. Rev. B}\ }\textbf {\bibinfo {volume} {103}},\ \bibinfo
  {pages} {174417} (\bibinfo {year} {2021})}\BibitemShut {NoStop}%
\bibitem [{\citenamefont {Suzuki}\ \emph {et~al.}(2021)\citenamefont {Suzuki},
  \citenamefont {Liu}, \citenamefont {Bertinshaw}, \citenamefont {Ueda},
  \citenamefont {Kim}, \citenamefont {Laha}, \citenamefont {Weber},
  \citenamefont {Yang}, \citenamefont {Wang}, \citenamefont {Takahashi},
  \citenamefont {Fürsich}, \citenamefont {Minola}, \citenamefont {Lotsch},
  \citenamefont {Kim}, \citenamefont {Yavaş}, \citenamefont {Daghofer},
  \citenamefont {Chaloupka}, \citenamefont {Khaliullin}, \citenamefont
  {Gretarsson},\ and\ \citenamefont {Keimer}}]{suzukiproximate2021}%
  \BibitemOpen
  \bibfield  {author} {\bibinfo {author} {\bibfnamefont {H.}~\bibnamefont
  {Suzuki}}, \bibinfo {author} {\bibfnamefont {H.}~\bibnamefont {Liu}},
  \bibinfo {author} {\bibfnamefont {J.}~\bibnamefont {Bertinshaw}}, \bibinfo
  {author} {\bibfnamefont {K.}~\bibnamefont {Ueda}}, \bibinfo {author}
  {\bibfnamefont {H.}~\bibnamefont {Kim}}, \bibinfo {author} {\bibfnamefont
  {S.}~\bibnamefont {Laha}}, \bibinfo {author} {\bibfnamefont {D.}~\bibnamefont
  {Weber}}, \bibinfo {author} {\bibfnamefont {Z.}~\bibnamefont {Yang}},
  \bibinfo {author} {\bibfnamefont {L.}~\bibnamefont {Wang}}, \bibinfo {author}
  {\bibfnamefont {H.}~\bibnamefont {Takahashi}}, \bibinfo {author}
  {\bibfnamefont {K.}~\bibnamefont {Fürsich}}, \bibinfo {author}
  {\bibfnamefont {M.}~\bibnamefont {Minola}}, \bibinfo {author} {\bibfnamefont
  {B.~V.}\ \bibnamefont {Lotsch}}, \bibinfo {author} {\bibfnamefont {B.~J.}\
  \bibnamefont {Kim}}, \bibinfo {author} {\bibfnamefont {H.}~\bibnamefont
  {Yavaş}}, \bibinfo {author} {\bibfnamefont {M.}~\bibnamefont {Daghofer}},
  \bibinfo {author} {\bibfnamefont {J.}~\bibnamefont {Chaloupka}}, \bibinfo
  {author} {\bibfnamefont {G.}~\bibnamefont {Khaliullin}}, \bibinfo {author}
  {\bibfnamefont {H.}~\bibnamefont {Gretarsson}},\ and\ \bibinfo {author}
  {\bibfnamefont {B.}~\bibnamefont {Keimer}},\ }\bibfield  {title} {\bibinfo
  {title} {Proximate ferromagnetic state in the {Kitaev} model material
  $\alpha$-{RuCl3}},\ }\href {https://doi.org/10.1038/s41467-021-24722-4}
  {\bibfield  {journal} {\bibinfo  {journal} {Nat. Commun.}\ }\textbf {\bibinfo
  {volume} {12}},\ \bibinfo {pages} {4512} (\bibinfo {year}
  {2021})}\BibitemShut {NoStop}%
\bibitem [{\citenamefont {Wagner}\ \emph {et~al.}(2022)\citenamefont {Wagner},
  \citenamefont {Sahasrabudhe}, \citenamefont {Versteeg}, \citenamefont
  {Wysocki}, \citenamefont {Wang}, \citenamefont {Tsurkan}, \citenamefont
  {Loidl}, \citenamefont {Khomskii}, \citenamefont {Hedayat},\ and\
  \citenamefont {van Loosdrecht}}]{wagnermagnetooptical2022}%
  \BibitemOpen
  \bibfield  {author} {\bibinfo {author} {\bibfnamefont {J.}~\bibnamefont
  {Wagner}}, \bibinfo {author} {\bibfnamefont {A.}~\bibnamefont
  {Sahasrabudhe}}, \bibinfo {author} {\bibfnamefont {R.~B.}\ \bibnamefont
  {Versteeg}}, \bibinfo {author} {\bibfnamefont {L.}~\bibnamefont {Wysocki}},
  \bibinfo {author} {\bibfnamefont {Z.}~\bibnamefont {Wang}}, \bibinfo {author}
  {\bibfnamefont {V.}~\bibnamefont {Tsurkan}}, \bibinfo {author} {\bibfnamefont
  {A.}~\bibnamefont {Loidl}}, \bibinfo {author} {\bibfnamefont {D.~I.}\
  \bibnamefont {Khomskii}}, \bibinfo {author} {\bibfnamefont {H.}~\bibnamefont
  {Hedayat}},\ and\ \bibinfo {author} {\bibfnamefont {P.~H.~M.}\ \bibnamefont
  {van Loosdrecht}},\ }\bibfield  {title} {\bibinfo {title} {Magneto-optical
  study of metamagnetic transitions in the antiferromagnetic phase of
  $\alpha$-{RuCl3}},\ }\href {https://doi.org/10.1038/s41535-022-00434-w}
  {\bibfield  {journal} {\bibinfo  {journal} {npj Quantum Mater.}\ }\textbf
  {\bibinfo {volume} {7}},\ \bibinfo {pages} {28} (\bibinfo {year}
  {2022})}\BibitemShut {NoStop}%
\bibitem [{\citenamefont {Bachus}\ \emph {et~al.}(2020)\citenamefont {Bachus},
  \citenamefont {Kaib}, \citenamefont {Tokiwa}, \citenamefont {Jesche},
  \citenamefont {Tsurkan}, \citenamefont {Loidl}, \citenamefont {Winter},
  \citenamefont {Tsirlin}, \citenamefont {Valentí},\ and\ \citenamefont
  {Gegenwart}}]{bachusthermodynamic2020}%
  \BibitemOpen
  \bibfield  {author} {\bibinfo {author} {\bibfnamefont {S.}~\bibnamefont
  {Bachus}}, \bibinfo {author} {\bibfnamefont {D.~A.}\ \bibnamefont {Kaib}},
  \bibinfo {author} {\bibfnamefont {Y.}~\bibnamefont {Tokiwa}}, \bibinfo
  {author} {\bibfnamefont {A.}~\bibnamefont {Jesche}}, \bibinfo {author}
  {\bibfnamefont {V.}~\bibnamefont {Tsurkan}}, \bibinfo {author} {\bibfnamefont
  {A.}~\bibnamefont {Loidl}}, \bibinfo {author} {\bibfnamefont
  {S.}~\bibnamefont {Winter}}, \bibinfo {author} {\bibfnamefont
  {A.}~\bibnamefont {Tsirlin}}, \bibinfo {author} {\bibfnamefont
  {R.}~\bibnamefont {Valentí}},\ and\ \bibinfo {author} {\bibfnamefont
  {P.}~\bibnamefont {Gegenwart}},\ }\bibfield  {title} {\bibinfo {title}
  {Thermodynamic {Perspective} on {Field}-{Induced} {Behavior} of
  $\alpha$-{RuCl} 3},\ }\href {https://doi.org/10.1103/PhysRevLett.125.097203}
  {\bibfield  {journal} {\bibinfo  {journal} {Phys. Rev. Lett.}\ }\textbf
  {\bibinfo {volume} {125}},\ \bibinfo {pages} {097203} (\bibinfo {year}
  {2020})}\BibitemShut {NoStop}%
\bibitem [{\citenamefont {Banerjee}\ \emph {et~al.}(2017)\citenamefont
  {Banerjee}, \citenamefont {Yan}, \citenamefont {Knolle}, \citenamefont
  {Bridges}, \citenamefont {Stone}, \citenamefont {Lumsden}, \citenamefont
  {Mandrus}, \citenamefont {Tennant}, \citenamefont {Moessner},\ and\
  \citenamefont {Nagler}}]{banerjeeneutron2017}%
  \BibitemOpen
  \bibfield  {author} {\bibinfo {author} {\bibfnamefont {A.}~\bibnamefont
  {Banerjee}}, \bibinfo {author} {\bibfnamefont {J.}~\bibnamefont {Yan}},
  \bibinfo {author} {\bibfnamefont {J.}~\bibnamefont {Knolle}}, \bibinfo
  {author} {\bibfnamefont {C.~A.}\ \bibnamefont {Bridges}}, \bibinfo {author}
  {\bibfnamefont {M.~B.}\ \bibnamefont {Stone}}, \bibinfo {author}
  {\bibfnamefont {M.~D.}\ \bibnamefont {Lumsden}}, \bibinfo {author}
  {\bibfnamefont {D.~G.}\ \bibnamefont {Mandrus}}, \bibinfo {author}
  {\bibfnamefont {D.~A.}\ \bibnamefont {Tennant}}, \bibinfo {author}
  {\bibfnamefont {R.}~\bibnamefont {Moessner}},\ and\ \bibinfo {author}
  {\bibfnamefont {S.~E.}\ \bibnamefont {Nagler}},\ }\bibfield  {title}
  {\bibinfo {title} {Neutron scattering in the proximate quantum spin liquid
  $\alpha$-{RuCl} $_{\textrm{3}}$},\ }\href
  {https://doi.org/10.1126/science.aah6015} {\bibfield  {journal} {\bibinfo
  {journal} {Science}\ }\textbf {\bibinfo {volume} {356}},\ \bibinfo {pages}
  {1055} (\bibinfo {year} {2017})}\BibitemShut {NoStop}%
\bibitem [{\citenamefont {Wolter}\ and\ \citenamefont
  {Hess}(2022)}]{wolterspin2022}%
  \BibitemOpen
  \bibfield  {author} {\bibinfo {author} {\bibfnamefont {A.~U.~B.}\
  \bibnamefont {Wolter}}\ and\ \bibinfo {author} {\bibfnamefont
  {C.}~\bibnamefont {Hess}},\ }\bibfield  {title} {\bibinfo {title} {Spin
  liquid evidence at the edge and in bulk},\ }\href
  {https://doi.org/10.1038/s41567-022-01521-2} {\bibfield  {journal} {\bibinfo
  {journal} {Nat. Phys.}\ }\textbf {\bibinfo {volume} {18}},\ \bibinfo {pages}
  {378} (\bibinfo {year} {2022})}\BibitemShut {NoStop}%
\bibitem [{\citenamefont {Knolle}\ \emph
  {et~al.}(2014{\natexlab{a}})\citenamefont {Knolle}, \citenamefont
  {Kovrizhin}, \citenamefont {Chalker},\ and\ \citenamefont
  {Moessner}}]{knolle14}%
  \BibitemOpen
  \bibfield  {author} {\bibinfo {author} {\bibfnamefont {J.}~\bibnamefont
  {Knolle}}, \bibinfo {author} {\bibfnamefont {D.~L.}\ \bibnamefont
  {Kovrizhin}}, \bibinfo {author} {\bibfnamefont {J.~T.}\ \bibnamefont
  {Chalker}},\ and\ \bibinfo {author} {\bibfnamefont {R.}~\bibnamefont
  {Moessner}},\ }\bibfield  {title} {\bibinfo {title} {{Dynamics of a
  Two-Dimensional Quantum Spin Liquid: Signatures of Emergent Majorana Fermions
  and Fluxes}},\ }\href {https://doi.org/10.1103/PhysRevLett.112.207203}
  {\bibfield  {journal} {\bibinfo  {journal} {Phys. Rev. Lett.}\ }\textbf
  {\bibinfo {volume} {112}},\ \bibinfo {pages} {207203} (\bibinfo {year}
  {2014}{\natexlab{a}})}\BibitemShut {NoStop}%
\bibitem [{\citenamefont {Knolle}\ \emph
  {et~al.}(2014{\natexlab{b}})\citenamefont {Knolle}, \citenamefont {Chern},
  \citenamefont {Kovrizhin}, \citenamefont {Moessner},\ and\ \citenamefont
  {Perkins}}]{knolle2014raman}%
  \BibitemOpen
  \bibfield  {author} {\bibinfo {author} {\bibfnamefont {J.}~\bibnamefont
  {Knolle}}, \bibinfo {author} {\bibfnamefont {G.-W.}\ \bibnamefont {Chern}},
  \bibinfo {author} {\bibfnamefont {D.~L.}\ \bibnamefont {Kovrizhin}}, \bibinfo
  {author} {\bibfnamefont {R.}~\bibnamefont {Moessner}},\ and\ \bibinfo
  {author} {\bibfnamefont {N.~B.}\ \bibnamefont {Perkins}},\ }\bibfield
  {title} {\bibinfo {title} {Raman scattering signatures of kitaev spin liquids
  in ${A}_{2}{\mathrm{iro}}_{3}$ iridates with $a=\mathrm{Na}$ or li},\ }\href
  {https://doi.org/10.1103/PhysRevLett.113.187201} {\bibfield  {journal}
  {\bibinfo  {journal} {Phys. Rev. Lett.}\ }\textbf {\bibinfo {volume} {113}},\
  \bibinfo {pages} {187201} (\bibinfo {year} {2014}{\natexlab{b}})}\BibitemShut
  {NoStop}%
\bibitem [{\citenamefont {Perreault}\ \emph {et~al.}(2015)\citenamefont
  {Perreault}, \citenamefont {Knolle}, \citenamefont {Perkins},\ and\
  \citenamefont {Burnell}}]{perreault2015theory}%
  \BibitemOpen
  \bibfield  {author} {\bibinfo {author} {\bibfnamefont {B.}~\bibnamefont
  {Perreault}}, \bibinfo {author} {\bibfnamefont {J.}~\bibnamefont {Knolle}},
  \bibinfo {author} {\bibfnamefont {N.~B.}\ \bibnamefont {Perkins}},\ and\
  \bibinfo {author} {\bibfnamefont {F.~J.}\ \bibnamefont {Burnell}},\
  }\bibfield  {title} {\bibinfo {title} {Theory of raman response in
  three-dimensional kitaev spin liquids: Application to $\ensuremath{\beta}$-
  and $\ensuremath{\gamma}\ensuremath{-}{\mathrm{li}}_{2}{\mathrm{iro}}_{3}$
  compounds},\ }\href {https://doi.org/10.1103/PhysRevB.92.094439} {\bibfield
  {journal} {\bibinfo  {journal} {Phys. Rev. B}\ }\textbf {\bibinfo {volume}
  {92}},\ \bibinfo {pages} {094439} (\bibinfo {year} {2015})}\BibitemShut
  {NoStop}%
\bibitem [{\citenamefont {Perreault}\ \emph {et~al.}(2016)\citenamefont
  {Perreault}, \citenamefont {Knolle}, \citenamefont {Perkins},\ and\
  \citenamefont {Burnell}}]{perreault2016resonant}%
  \BibitemOpen
  \bibfield  {author} {\bibinfo {author} {\bibfnamefont {B.}~\bibnamefont
  {Perreault}}, \bibinfo {author} {\bibfnamefont {J.}~\bibnamefont {Knolle}},
  \bibinfo {author} {\bibfnamefont {N.~B.}\ \bibnamefont {Perkins}},\ and\
  \bibinfo {author} {\bibfnamefont {F.~J.}\ \bibnamefont {Burnell}},\
  }\bibfield  {title} {\bibinfo {title} {Resonant raman scattering theory for
  kitaev models and their majorana fermion boundary modes},\ }\href
  {https://doi.org/10.1103/PhysRevB.94.104427} {\bibfield  {journal} {\bibinfo
  {journal} {Phys. Rev. B}\ }\textbf {\bibinfo {volume} {94}},\ \bibinfo
  {pages} {104427} (\bibinfo {year} {2016})}\BibitemShut {NoStop}%
\bibitem [{\citenamefont {Rousochatzakis}\ \emph {et~al.}(2019)\citenamefont
  {Rousochatzakis}, \citenamefont {Kourtis}, \citenamefont {Knolle},
  \citenamefont {Moessner},\ and\ \citenamefont
  {Perkins}}]{rousochatzakis2019quantum}%
  \BibitemOpen
  \bibfield  {author} {\bibinfo {author} {\bibfnamefont {I.}~\bibnamefont
  {Rousochatzakis}}, \bibinfo {author} {\bibfnamefont {S.}~\bibnamefont
  {Kourtis}}, \bibinfo {author} {\bibfnamefont {J.}~\bibnamefont {Knolle}},
  \bibinfo {author} {\bibfnamefont {R.}~\bibnamefont {Moessner}},\ and\
  \bibinfo {author} {\bibfnamefont {N.~B.}\ \bibnamefont {Perkins}},\
  }\bibfield  {title} {\bibinfo {title} {Quantum spin liquid at finite
  temperature: Proximate dynamics and persistent typicality},\ }\href
  {https://doi.org/10.1103/PhysRevB.100.045117} {\bibfield  {journal} {\bibinfo
   {journal} {Phys. Rev. B}\ }\textbf {\bibinfo {volume} {100}},\ \bibinfo
  {pages} {045117} (\bibinfo {year} {2019})}\BibitemShut {NoStop}%
\bibitem [{\citenamefont {Hal\'asz}\ \emph {et~al.}(2016)\citenamefont
  {Hal\'asz}, \citenamefont {Perkins},\ and\ \citenamefont {van~den
  Brink}}]{Gabor2016}%
  \BibitemOpen
  \bibfield  {author} {\bibinfo {author} {\bibfnamefont {G.~B.}\ \bibnamefont
  {Hal\'asz}}, \bibinfo {author} {\bibfnamefont {N.~B.}\ \bibnamefont
  {Perkins}},\ and\ \bibinfo {author} {\bibfnamefont {J.}~\bibnamefont {van~den
  Brink}},\ }\bibfield  {title} {\bibinfo {title} {{Resonant Inelastic X-Ray
  Scattering Response of the Kitaev Honeycomb Model}},\ }\href
  {https://doi.org/10.1103/PhysRevLett.117.127203} {\bibfield  {journal}
  {\bibinfo  {journal} {Phys. Rev. Lett.}\ }\textbf {\bibinfo {volume} {117}},\
  \bibinfo {pages} {127203} (\bibinfo {year} {2016})}\BibitemShut {NoStop}%
\bibitem [{\citenamefont {Hal\'asz}\ \emph {et~al.}(2019)\citenamefont
  {Hal\'asz}, \citenamefont {Kourtis}, \citenamefont {Knolle},\ and\
  \citenamefont {Perkins}}]{halasz2019observing}%
  \BibitemOpen
  \bibfield  {author} {\bibinfo {author} {\bibfnamefont {G.~B.}\ \bibnamefont
  {Hal\'asz}}, \bibinfo {author} {\bibfnamefont {S.}~\bibnamefont {Kourtis}},
  \bibinfo {author} {\bibfnamefont {J.}~\bibnamefont {Knolle}},\ and\ \bibinfo
  {author} {\bibfnamefont {N.~B.}\ \bibnamefont {Perkins}},\ }\bibfield
  {title} {\bibinfo {title} {Observing spin fractionalization in the kitaev
  spin liquid via temperature evolution of indirect resonant inelastic x-ray
  scattering},\ }\href {https://doi.org/10.1103/PhysRevB.99.184417} {\bibfield
  {journal} {\bibinfo  {journal} {Phys. Rev. B}\ }\textbf {\bibinfo {volume}
  {99}},\ \bibinfo {pages} {184417} (\bibinfo {year} {2019})}\BibitemShut
  {NoStop}%
\bibitem [{\citenamefont {Wan}\ and\ \citenamefont
  {Armitage}(2019)}]{wan2019resolving}%
  \BibitemOpen
  \bibfield  {author} {\bibinfo {author} {\bibfnamefont {Y.}~\bibnamefont
  {Wan}}\ and\ \bibinfo {author} {\bibfnamefont {N.~P.}\ \bibnamefont
  {Armitage}},\ }\bibfield  {title} {\bibinfo {title} {Resolving continua of
  fractional excitations by spinon echo in thz 2d coherent spectroscopy},\
  }\href {https://doi.org/10.1103/PhysRevLett.122.257401} {\bibfield  {journal}
  {\bibinfo  {journal} {Phys. Rev. Lett.}\ }\textbf {\bibinfo {volume} {122}},\
  \bibinfo {pages} {257401} (\bibinfo {year} {2019})}\BibitemShut {NoStop}%
\bibitem [{\citenamefont {Pereira}\ and\ \citenamefont
  {Egger}(2020)}]{pereira2020electrical}%
  \BibitemOpen
  \bibfield  {author} {\bibinfo {author} {\bibfnamefont {R.~G.}\ \bibnamefont
  {Pereira}}\ and\ \bibinfo {author} {\bibfnamefont {R.}~\bibnamefont
  {Egger}},\ }\bibfield  {title} {\bibinfo {title} {Electrical access to ising
  anyons in kitaev spin liquids},\ }\href
  {https://doi.org/10.1103/PhysRevLett.125.227202} {\bibfield  {journal}
  {\bibinfo  {journal} {Phys. Rev. Lett.}\ }\textbf {\bibinfo {volume} {125}},\
  \bibinfo {pages} {227202} (\bibinfo {year} {2020})}\BibitemShut {NoStop}%
\bibitem [{\citenamefont {Udagawa}\ \emph {et~al.}(2021)\citenamefont
  {Udagawa}, \citenamefont {Takayoshi},\ and\ \citenamefont
  {Oka}}]{udagawa2021scanning}%
  \BibitemOpen
  \bibfield  {author} {\bibinfo {author} {\bibfnamefont {M.}~\bibnamefont
  {Udagawa}}, \bibinfo {author} {\bibfnamefont {S.}~\bibnamefont {Takayoshi}},\
  and\ \bibinfo {author} {\bibfnamefont {T.}~\bibnamefont {Oka}},\ }\bibfield
  {title} {\bibinfo {title} {Scanning tunneling microscopy as a single majorana
  detector of kitaev's chiral spin liquid},\ }\href
  {https://doi.org/10.1103/PhysRevLett.126.127201} {\bibfield  {journal}
  {\bibinfo  {journal} {Phys. Rev. Lett.}\ }\textbf {\bibinfo {volume} {126}},\
  \bibinfo {pages} {127201} (\bibinfo {year} {2021})}\BibitemShut {NoStop}%
\bibitem [{\citenamefont {Joy}\ and\ \citenamefont
  {Rosch}(2022)}]{joy2022dynamics}%
  \BibitemOpen
  \bibfield  {author} {\bibinfo {author} {\bibfnamefont {A.~P.}\ \bibnamefont
  {Joy}}\ and\ \bibinfo {author} {\bibfnamefont {A.}~\bibnamefont {Rosch}},\
  }\bibfield  {title} {\bibinfo {title} {Dynamics of visons and thermal hall
  effect in perturbed kitaev models},\ }\href
  {https://doi.org/10.1103/PhysRevX.12.041004} {\bibfield  {journal} {\bibinfo
  {journal} {Phys. Rev. X}\ }\textbf {\bibinfo {volume} {12}},\ \bibinfo
  {pages} {041004} (\bibinfo {year} {2022})}\BibitemShut {NoStop}%
\bibitem [{\citenamefont {Bauer}\ \emph {et~al.}(2023)\citenamefont {Bauer},
  \citenamefont {Freitas}, \citenamefont {Pereira},\ and\ \citenamefont
  {Egger}}]{bauer2023scanning}%
  \BibitemOpen
  \bibfield  {author} {\bibinfo {author} {\bibfnamefont {T.}~\bibnamefont
  {Bauer}}, \bibinfo {author} {\bibfnamefont {L.~R.~D.}\ \bibnamefont
  {Freitas}}, \bibinfo {author} {\bibfnamefont {R.~G.}\ \bibnamefont
  {Pereira}},\ and\ \bibinfo {author} {\bibfnamefont {R.}~\bibnamefont
  {Egger}},\ }\bibfield  {title} {\bibinfo {title} {Scanning tunneling
  spectroscopy of majorana zero modes in a kitaev spin liquid},\ }\href
  {https://doi.org/10.1103/PhysRevB.107.054432} {\bibfield  {journal} {\bibinfo
   {journal} {Phys. Rev. B}\ }\textbf {\bibinfo {volume} {107}},\ \bibinfo
  {pages} {054432} (\bibinfo {year} {2023})}\BibitemShut {NoStop}%
\bibitem [{\citenamefont {Kao}\ \emph {et~al.}(2024{\natexlab{a}})\citenamefont
  {Kao}, \citenamefont {Hal\'asz},\ and\ \citenamefont
  {Perkins}}]{kao2024dynamics}%
  \BibitemOpen
  \bibfield  {author} {\bibinfo {author} {\bibfnamefont {W.-H.}\ \bibnamefont
  {Kao}}, \bibinfo {author} {\bibfnamefont {G.~B.}\ \bibnamefont {Hal\'asz}},\
  and\ \bibinfo {author} {\bibfnamefont {N.~B.}\ \bibnamefont {Perkins}},\
  }\bibfield  {title} {\bibinfo {title} {Dynamics of vacancy-induced modes in
  the non-abelian kitaev spin liquid},\ }\href
  {https://doi.org/10.1103/PhysRevB.109.125150} {\bibfield  {journal} {\bibinfo
   {journal} {Phys. Rev. B}\ }\textbf {\bibinfo {volume} {109}},\ \bibinfo
  {pages} {125150} (\bibinfo {year} {2024}{\natexlab{a}})}\BibitemShut
  {NoStop}%
\bibitem [{\citenamefont {Kao}\ \emph {et~al.}(2024{\natexlab{b}})\citenamefont
  {Kao}, \citenamefont {Perkins},\ and\ \citenamefont {Hal\'asz}}]{kao2024STM}%
  \BibitemOpen
  \bibfield  {author} {\bibinfo {author} {\bibfnamefont {W.-H.}\ \bibnamefont
  {Kao}}, \bibinfo {author} {\bibfnamefont {N.~B.}\ \bibnamefont {Perkins}},\
  and\ \bibinfo {author} {\bibfnamefont {G.~B.}\ \bibnamefont {Hal\'asz}},\
  }\bibfield  {title} {\bibinfo {title} {Vacancy spectroscopy of non-abelian
  kitaev spin liquids},\ }\href
  {https://doi.org/10.1103/PhysRevLett.132.136503} {\bibfield  {journal}
  {\bibinfo  {journal} {Phys. Rev. Lett.}\ }\textbf {\bibinfo {volume} {132}},\
  \bibinfo {pages} {136503} (\bibinfo {year} {2024}{\natexlab{b}})}\BibitemShut
  {NoStop}%
\bibitem [{\citenamefont {Vinkler-Aviv}\ and\ \citenamefont
  {Rosch}(2018)}]{aviv18}%
  \BibitemOpen
  \bibfield  {author} {\bibinfo {author} {\bibfnamefont {Y.}~\bibnamefont
  {Vinkler-Aviv}}\ and\ \bibinfo {author} {\bibfnamefont {A.}~\bibnamefont
  {Rosch}},\ }\bibfield  {title} {\bibinfo {title} {{Approximately Quantized
  Thermal Hall Effect of Chiral Liquids Coupled to Phonons}},\ }\href
  {https://doi.org/10.1103/PhysRevX.8.031032} {\bibfield  {journal} {\bibinfo
  {journal} {Phys. Rev. X}\ }\textbf {\bibinfo {volume} {8}},\ \bibinfo {pages}
  {031032} (\bibinfo {year} {2018})}\BibitemShut {NoStop}%
\bibitem [{\citenamefont {Ye}\ \emph {et~al.}(2018)\citenamefont {Ye},
  \citenamefont {Hal\'asz}, \citenamefont {Savary},\ and\ \citenamefont
  {Balents}}]{ye18}%
  \BibitemOpen
  \bibfield  {author} {\bibinfo {author} {\bibfnamefont {M.}~\bibnamefont
  {Ye}}, \bibinfo {author} {\bibfnamefont {G.~B.}\ \bibnamefont {Hal\'asz}},
  \bibinfo {author} {\bibfnamefont {L.}~\bibnamefont {Savary}},\ and\ \bibinfo
  {author} {\bibfnamefont {L.}~\bibnamefont {Balents}},\ }\bibfield  {title}
  {\bibinfo {title} {{Quantization of the Thermal Hall Conductivity at Small
  Hall Angles}},\ }\href {https://doi.org/10.1103/PhysRevLett.121.147201}
  {\bibfield  {journal} {\bibinfo  {journal} {Phys. Rev. Lett.}\ }\textbf
  {\bibinfo {volume} {121}},\ \bibinfo {pages} {147201} (\bibinfo {year}
  {2018})}\BibitemShut {NoStop}%
\bibitem [{\citenamefont {Metavitsiadis}\ and\ \citenamefont
  {Brenig}(2020)}]{metavitsiadisphonon2020}%
  \BibitemOpen
  \bibfield  {author} {\bibinfo {author} {\bibfnamefont {A.}~\bibnamefont
  {Metavitsiadis}}\ and\ \bibinfo {author} {\bibfnamefont {W.}~\bibnamefont
  {Brenig}},\ }\bibfield  {title} {\bibinfo {title} {Phonon renormalization in
  the kitaev quantum spin liquid},\ }\href
  {https://doi.org/10.1103/PhysRevB.101.035103} {\bibfield  {journal} {\bibinfo
   {journal} {Phys. Rev. B}\ }\textbf {\bibinfo {volume} {101}},\ \bibinfo
  {pages} {035103} (\bibinfo {year} {2020})}\BibitemShut {NoStop}%
\bibitem [{\citenamefont {Ye}\ \emph {et~al.}(2020)\citenamefont {Ye},
  \citenamefont {Fernandes},\ and\ \citenamefont {Perkins}}]{yephonon2020}%
  \BibitemOpen
  \bibfield  {author} {\bibinfo {author} {\bibfnamefont {M.}~\bibnamefont
  {Ye}}, \bibinfo {author} {\bibfnamefont {R.~M.}\ \bibnamefont {Fernandes}},\
  and\ \bibinfo {author} {\bibfnamefont {N.~B.}\ \bibnamefont {Perkins}},\
  }\bibfield  {title} {\bibinfo {title} {Phonon dynamics in the kitaev spin
  liquid},\ }\href {https://doi.org/10.1103/PhysRevResearch.2.033180}
  {\bibfield  {journal} {\bibinfo  {journal} {Phys. Rev. Research}\ }\textbf
  {\bibinfo {volume} {2}},\ \bibinfo {pages} {033180} (\bibinfo {year}
  {2020})}\BibitemShut {NoStop}%
\bibitem [{\citenamefont {Li}\ \emph {et~al.}(2021{\natexlab{a}})\citenamefont
  {Li}, \citenamefont {Zhang}, \citenamefont {Said}, \citenamefont {Fabbris},
  \citenamefont {Mazzone}, \citenamefont {Yan}, \citenamefont {Mandrus},
  \citenamefont {Halász}, \citenamefont {Okamoto}, \citenamefont {Murakami},
  \citenamefont {Dean}, \citenamefont {Lee},\ and\ \citenamefont
  {Miao}}]{ligiant2021}%
  \BibitemOpen
  \bibfield  {author} {\bibinfo {author} {\bibfnamefont {H.}~\bibnamefont
  {Li}}, \bibinfo {author} {\bibfnamefont {T.~T.}\ \bibnamefont {Zhang}},
  \bibinfo {author} {\bibfnamefont {A.}~\bibnamefont {Said}}, \bibinfo {author}
  {\bibfnamefont {G.}~\bibnamefont {Fabbris}}, \bibinfo {author} {\bibfnamefont
  {D.~G.}\ \bibnamefont {Mazzone}}, \bibinfo {author} {\bibfnamefont {J.~Q.}\
  \bibnamefont {Yan}}, \bibinfo {author} {\bibfnamefont {D.}~\bibnamefont
  {Mandrus}}, \bibinfo {author} {\bibfnamefont {G.~B.}\ \bibnamefont
  {Halász}}, \bibinfo {author} {\bibfnamefont {S.}~\bibnamefont {Okamoto}},
  \bibinfo {author} {\bibfnamefont {S.}~\bibnamefont {Murakami}}, \bibinfo
  {author} {\bibfnamefont {M.~P.~M.}\ \bibnamefont {Dean}}, \bibinfo {author}
  {\bibfnamefont {H.~N.}\ \bibnamefont {Lee}},\ and\ \bibinfo {author}
  {\bibfnamefont {H.}~\bibnamefont {Miao}},\ }\bibfield  {title} {\bibinfo
  {title} {Giant phonon anomalies in the proximate {Kitaev} quantum spin liquid
  $\alpha$-{RuCl3}},\ }\href {https://doi.org/10.1038/s41467-021-23826-1}
  {\bibfield  {journal} {\bibinfo  {journal} {Nat. Commun.}\ }\textbf {\bibinfo
  {volume} {12}},\ \bibinfo {pages} {3513} (\bibinfo {year}
  {2021}{\natexlab{a}})}\BibitemShut {NoStop}%
\bibitem [{\citenamefont {Kocsis}\ \emph {et~al.}(2022)\citenamefont {Kocsis},
  \citenamefont {Kaib}, \citenamefont {Riedl}, \citenamefont {Gass},
  \citenamefont {Lampen-Kelley}, \citenamefont {Mandrus}, \citenamefont
  {Nagler}, \citenamefont {P\'erez}, \citenamefont {Nielsch}, \citenamefont
  {B\"uchner}, \citenamefont {Wolter},\ and\ \citenamefont
  {Valent\'{\i}}}]{kocsismagnetoelastic2022}%
  \BibitemOpen
  \bibfield  {author} {\bibinfo {author} {\bibfnamefont {V.}~\bibnamefont
  {Kocsis}}, \bibinfo {author} {\bibfnamefont {D.~A.~S.}\ \bibnamefont {Kaib}},
  \bibinfo {author} {\bibfnamefont {K.}~\bibnamefont {Riedl}}, \bibinfo
  {author} {\bibfnamefont {S.}~\bibnamefont {Gass}}, \bibinfo {author}
  {\bibfnamefont {P.}~\bibnamefont {Lampen-Kelley}}, \bibinfo {author}
  {\bibfnamefont {D.~G.}\ \bibnamefont {Mandrus}}, \bibinfo {author}
  {\bibfnamefont {S.~E.}\ \bibnamefont {Nagler}}, \bibinfo {author}
  {\bibfnamefont {N.}~\bibnamefont {P\'erez}}, \bibinfo {author} {\bibfnamefont
  {K.}~\bibnamefont {Nielsch}}, \bibinfo {author} {\bibfnamefont
  {B.}~\bibnamefont {B\"uchner}}, \bibinfo {author} {\bibfnamefont {A.~U.~B.}\
  \bibnamefont {Wolter}},\ and\ \bibinfo {author} {\bibfnamefont
  {R.}~\bibnamefont {Valent\'{\i}}},\ }\bibfield  {title} {\bibinfo {title}
  {Magnetoelastic coupling anisotropy in the kitaev material
  $\ensuremath{\alpha}\text{\ensuremath{-}}\mathrm{Ru}{\mathrm{cl}}_{3}$},\
  }\href {https://doi.org/10.1103/PhysRevB.105.094410} {\bibfield  {journal}
  {\bibinfo  {journal} {Phys. Rev. B}\ }\textbf {\bibinfo {volume} {105}},\
  \bibinfo {pages} {094410} (\bibinfo {year} {2022})}\BibitemShut {NoStop}%
\bibitem [{\citenamefont {Kaib}\ \emph
  {et~al.}(2021{\natexlab{a}})\citenamefont {Kaib}, \citenamefont {Biswas},
  \citenamefont {Riedl}, \citenamefont {Winter},\ and\ \citenamefont
  {Valent\'{\i}}}]{kaibmagnetoelastic2021}%
  \BibitemOpen
  \bibfield  {author} {\bibinfo {author} {\bibfnamefont {D.~A.~S.}\
  \bibnamefont {Kaib}}, \bibinfo {author} {\bibfnamefont {S.}~\bibnamefont
  {Biswas}}, \bibinfo {author} {\bibfnamefont {K.}~\bibnamefont {Riedl}},
  \bibinfo {author} {\bibfnamefont {S.~M.}\ \bibnamefont {Winter}},\ and\
  \bibinfo {author} {\bibfnamefont {R.}~\bibnamefont {Valent\'{\i}}},\
  }\bibfield  {title} {\bibinfo {title} {Magnetoelastic coupling and effects of
  uniaxial strain in $\ensuremath{\alpha}\ensuremath{-}{\mathrm{rucl}}_{3}$
  from first principles},\ }\href
  {https://doi.org/10.1103/PhysRevB.103.L140402} {\bibfield  {journal}
  {\bibinfo  {journal} {Phys. Rev. B}\ }\textbf {\bibinfo {volume} {103}},\
  \bibinfo {pages} {L140402} (\bibinfo {year}
  {2021}{\natexlab{a}})}\BibitemShut {NoStop}%
\bibitem [{\citenamefont {Sch\"onemann}\ \emph {et~al.}(2020)\citenamefont
  {Sch\"onemann}, \citenamefont {Imajo}, \citenamefont {Weickert},
  \citenamefont {Yan}, \citenamefont {Mandrus}, \citenamefont {Takano},
  \citenamefont {Brosha}, \citenamefont {Rosa}, \citenamefont {Nagler},
  \citenamefont {Kindo},\ and\ \citenamefont {Jaime}}]{schoenemannthermal2020}%
  \BibitemOpen
  \bibfield  {author} {\bibinfo {author} {\bibfnamefont {R.}~\bibnamefont
  {Sch\"onemann}}, \bibinfo {author} {\bibfnamefont {S.}~\bibnamefont {Imajo}},
  \bibinfo {author} {\bibfnamefont {F.}~\bibnamefont {Weickert}}, \bibinfo
  {author} {\bibfnamefont {J.}~\bibnamefont {Yan}}, \bibinfo {author}
  {\bibfnamefont {D.~G.}\ \bibnamefont {Mandrus}}, \bibinfo {author}
  {\bibfnamefont {Y.}~\bibnamefont {Takano}}, \bibinfo {author} {\bibfnamefont
  {E.~L.}\ \bibnamefont {Brosha}}, \bibinfo {author} {\bibfnamefont {P.~F.~S.}\
  \bibnamefont {Rosa}}, \bibinfo {author} {\bibfnamefont {S.~E.}\ \bibnamefont
  {Nagler}}, \bibinfo {author} {\bibfnamefont {K.}~\bibnamefont {Kindo}},\ and\
  \bibinfo {author} {\bibfnamefont {M.}~\bibnamefont {Jaime}},\ }\bibfield
  {title} {\bibinfo {title} {Thermal and magnetoelastic properties of
  $\ensuremath{\alpha}\ensuremath{-}{\mathrm{rucl}}_{3}$ in the field-induced
  low-temperature states},\ }\href
  {https://doi.org/10.1103/PhysRevB.102.214432} {\bibfield  {journal} {\bibinfo
   {journal} {Phys. Rev. B}\ }\textbf {\bibinfo {volume} {102}},\ \bibinfo
  {pages} {214432} (\bibinfo {year} {2020})}\BibitemShut {NoStop}%
\bibitem [{\citenamefont {Hentrich}\ \emph {et~al.}(2018)\citenamefont
  {Hentrich}, \citenamefont {Wolter}, \citenamefont {Zotos}, \citenamefont
  {Brenig}, \citenamefont {Nowak}, \citenamefont {Isaeva}, \citenamefont
  {Doert}, \citenamefont {Banerjee}, \citenamefont {Lampen-Kelley},
  \citenamefont {Mandrus}, \citenamefont {Nagler}, \citenamefont {Sears},
  \citenamefont {Kim}, \citenamefont {B\"uchner},\ and\ \citenamefont
  {Hess}}]{hentrichunusual2018}%
  \BibitemOpen
  \bibfield  {author} {\bibinfo {author} {\bibfnamefont {R.}~\bibnamefont
  {Hentrich}}, \bibinfo {author} {\bibfnamefont {A.~U.~B.}\ \bibnamefont
  {Wolter}}, \bibinfo {author} {\bibfnamefont {X.}~\bibnamefont {Zotos}},
  \bibinfo {author} {\bibfnamefont {W.}~\bibnamefont {Brenig}}, \bibinfo
  {author} {\bibfnamefont {D.}~\bibnamefont {Nowak}}, \bibinfo {author}
  {\bibfnamefont {A.}~\bibnamefont {Isaeva}}, \bibinfo {author} {\bibfnamefont
  {T.}~\bibnamefont {Doert}}, \bibinfo {author} {\bibfnamefont
  {A.}~\bibnamefont {Banerjee}}, \bibinfo {author} {\bibfnamefont
  {P.}~\bibnamefont {Lampen-Kelley}}, \bibinfo {author} {\bibfnamefont {D.~G.}\
  \bibnamefont {Mandrus}}, \bibinfo {author} {\bibfnamefont {S.~E.}\
  \bibnamefont {Nagler}}, \bibinfo {author} {\bibfnamefont {J.}~\bibnamefont
  {Sears}}, \bibinfo {author} {\bibfnamefont {Y.-J.}\ \bibnamefont {Kim}},
  \bibinfo {author} {\bibfnamefont {B.}~\bibnamefont {B\"uchner}},\ and\
  \bibinfo {author} {\bibfnamefont {C.}~\bibnamefont {Hess}},\ }\bibfield
  {title} {\bibinfo {title} {Unusual phonon heat transport in
  $\ensuremath{\alpha}\text{\ensuremath{-}}{\mathrm{rucl}}_{3}$: Strong
  spin-phonon scattering and field-induced spin gap},\ }\href
  {https://doi.org/10.1103/PhysRevLett.120.117204} {\bibfield  {journal}
  {\bibinfo  {journal} {Phys. Rev. Lett.}\ }\textbf {\bibinfo {volume} {120}},\
  \bibinfo {pages} {117204} (\bibinfo {year} {2018})}\BibitemShut {NoStop}%
\bibitem [{\citenamefont {Reschke}\ \emph {et~al.}(2019)\citenamefont
  {Reschke}, \citenamefont {Tsurkan}, \citenamefont {Do}, \citenamefont {Choi},
  \citenamefont {Lunkenheimer}, \citenamefont {Wang},\ and\ \citenamefont
  {Loidl}}]{reschketerahertz2019}%
  \BibitemOpen
  \bibfield  {author} {\bibinfo {author} {\bibfnamefont {S.}~\bibnamefont
  {Reschke}}, \bibinfo {author} {\bibfnamefont {V.}~\bibnamefont {Tsurkan}},
  \bibinfo {author} {\bibfnamefont {S.-H.}\ \bibnamefont {Do}}, \bibinfo
  {author} {\bibfnamefont {K.-Y.}\ \bibnamefont {Choi}}, \bibinfo {author}
  {\bibfnamefont {P.}~\bibnamefont {Lunkenheimer}}, \bibinfo {author}
  {\bibfnamefont {Z.}~\bibnamefont {Wang}},\ and\ \bibinfo {author}
  {\bibfnamefont {A.}~\bibnamefont {Loidl}},\ }\bibfield  {title} {\bibinfo
  {title} {Terahertz excitations in
  $\ensuremath{\alpha}\text{\ensuremath{-}}\mathrm{RuC}{\mathrm{l}}_{3}$:
  Majorana fermions and rigid-plane shear and compression modes},\ }\href
  {https://doi.org/10.1103/PhysRevB.100.100403} {\bibfield  {journal} {\bibinfo
   {journal} {Phys. Rev. B}\ }\textbf {\bibinfo {volume} {100}},\ \bibinfo
  {pages} {100403} (\bibinfo {year} {2019})}\BibitemShut {NoStop}%
\bibitem [{\citenamefont {Bruin}\ \emph {et~al.}(2022)\citenamefont {Bruin},
  \citenamefont {Claus}, \citenamefont {Matsumoto}, \citenamefont {Kurita},
  \citenamefont {Tanaka},\ and\ \citenamefont {Takagi}}]{bruinrobustness2022}%
  \BibitemOpen
  \bibfield  {author} {\bibinfo {author} {\bibfnamefont {J.~A.~N.}\
  \bibnamefont {Bruin}}, \bibinfo {author} {\bibfnamefont {R.~R.}\ \bibnamefont
  {Claus}}, \bibinfo {author} {\bibfnamefont {Y.}~\bibnamefont {Matsumoto}},
  \bibinfo {author} {\bibfnamefont {N.}~\bibnamefont {Kurita}}, \bibinfo
  {author} {\bibfnamefont {H.}~\bibnamefont {Tanaka}},\ and\ \bibinfo {author}
  {\bibfnamefont {H.}~\bibnamefont {Takagi}},\ }\bibfield  {title} {\bibinfo
  {title} {Robustness of the thermal {Hall} effect close to half-quantization
  in $\alpha$-{RuCl3}},\ }\href {https://doi.org/10.1038/s41567-021-01501-y}
  {\bibfield  {journal} {\bibinfo  {journal} {Nat. Phys.}\ }\textbf {\bibinfo
  {volume} {18}},\ \bibinfo {pages} {401–} (\bibinfo {year}
  {2022})}\BibitemShut {NoStop}%
\bibitem [{\citenamefont {Yamashita}\ \emph {et~al.}(2020)\citenamefont
  {Yamashita}, \citenamefont {Gouchi}, \citenamefont {Uwatoko}, \citenamefont
  {Kurita},\ and\ \citenamefont {Tanaka}}]{yamashitasample2020}%
  \BibitemOpen
  \bibfield  {author} {\bibinfo {author} {\bibfnamefont {M.}~\bibnamefont
  {Yamashita}}, \bibinfo {author} {\bibfnamefont {J.}~\bibnamefont {Gouchi}},
  \bibinfo {author} {\bibfnamefont {Y.}~\bibnamefont {Uwatoko}}, \bibinfo
  {author} {\bibfnamefont {N.}~\bibnamefont {Kurita}},\ and\ \bibinfo {author}
  {\bibfnamefont {H.}~\bibnamefont {Tanaka}},\ }\bibfield  {title} {\bibinfo
  {title} {Sample dependence of half-integer quantized thermal hall effect in
  the kitaev spin-liquid candidate
  $\ensuremath{\alpha}\text{\ensuremath{-}}{\mathrm{rucl}}_{3}$},\ }\href
  {https://doi.org/10.1103/PhysRevB.102.220404} {\bibfield  {journal} {\bibinfo
   {journal} {Phys. Rev. B}\ }\textbf {\bibinfo {volume} {102}},\ \bibinfo
  {pages} {220404} (\bibinfo {year} {2020})}\BibitemShut {NoStop}%
\bibitem [{\citenamefont {Truell}\ \emph {et~al.}(1969)\citenamefont {Truell},
  \citenamefont {Elbaum},\ and\ \citenamefont {Chick}}]{truellultrasonic1969}%
  \BibitemOpen
  \bibfield  {author} {\bibinfo {author} {\bibfnamefont {R.}~\bibnamefont
  {Truell}}, \bibinfo {author} {\bibfnamefont {C.}~\bibnamefont {Elbaum}},\
  and\ \bibinfo {author} {\bibfnamefont {B.~B.}\ \bibnamefont {Chick}},\
  }\href@noop {} {\emph {\bibinfo {title} {Ultrasonic {Methods} in {Solid}
  {State} {Physics}}}}\ (\bibinfo  {publisher} {Academic Press, Inc.},\
  \bibinfo {address} {New York},\ \bibinfo {year} {1969})\BibitemShut {NoStop}%
\bibitem [{\citenamefont {Lüthi}(2005)}]{luthiphysical2005}%
  \BibitemOpen
  \bibfield  {author} {\bibinfo {author} {\bibfnamefont {B.}~\bibnamefont
  {Lüthi}},\ }\href@noop {} {\emph {\bibinfo {title} {Physical {Acoustics} in
  the {Solid} {State}}}}\ (\bibinfo  {publisher} {Springer-Verlag},\ \bibinfo
  {address} {Berlin Heidelberg, New York},\ \bibinfo {year} {2005})\BibitemShut
  {NoStop}%
\bibitem [{\citenamefont {Hauspurg}\ \emph {et~al.}(2024)\citenamefont
  {Hauspurg}, \citenamefont {Zherlitsyn}, \citenamefont {Helm}, \citenamefont
  {Felea}, \citenamefont {Wosnitza}, \citenamefont {Tsurkan}, \citenamefont
  {Choi}, \citenamefont {Do}, \citenamefont {Ye}, \citenamefont {Brenig},\ and\
  \citenamefont {Perkins}}]{hau24}%
  \BibitemOpen
  \bibfield  {author} {\bibinfo {author} {\bibfnamefont {A.}~\bibnamefont
  {Hauspurg}}, \bibinfo {author} {\bibfnamefont {S.}~\bibnamefont
  {Zherlitsyn}}, \bibinfo {author} {\bibfnamefont {T.}~\bibnamefont {Helm}},
  \bibinfo {author} {\bibfnamefont {V.}~\bibnamefont {Felea}}, \bibinfo
  {author} {\bibfnamefont {J.}~\bibnamefont {Wosnitza}}, \bibinfo {author}
  {\bibfnamefont {V.}~\bibnamefont {Tsurkan}}, \bibinfo {author} {\bibfnamefont
  {K.-Y.}\ \bibnamefont {Choi}}, \bibinfo {author} {\bibfnamefont {S.-H.}\
  \bibnamefont {Do}}, \bibinfo {author} {\bibfnamefont {M.}~\bibnamefont {Ye}},
  \bibinfo {author} {\bibfnamefont {W.}~\bibnamefont {Brenig}},\ and\ \bibinfo
  {author} {\bibfnamefont {N.~B.}\ \bibnamefont {Perkins}},\ }\bibfield
  {title} {\bibinfo {title} {Fractionalized excitations probed by ultrasound},\
  }\href {https://doi.org/10.1103/PhysRevB.109.144415} {\bibfield  {journal}
  {\bibinfo  {journal} {Phys. Rev. B}\ }\textbf {\bibinfo {volume} {109}},\
  \bibinfo {pages} {144415} (\bibinfo {year} {2024})}\BibitemShut {NoStop}%
\bibitem [{\citenamefont {Feng}\ \emph {et~al.}(2021)\citenamefont {Feng},
  \citenamefont {Ye},\ and\ \citenamefont {Perkins}}]{fengtemperature2021}%
  \BibitemOpen
  \bibfield  {author} {\bibinfo {author} {\bibfnamefont {K.}~\bibnamefont
  {Feng}}, \bibinfo {author} {\bibfnamefont {M.}~\bibnamefont {Ye}},\ and\
  \bibinfo {author} {\bibfnamefont {N.~B.}\ \bibnamefont {Perkins}},\
  }\bibfield  {title} {\bibinfo {title} {Temperature evolution of the phonon
  dynamics in the kitaev spin liquid},\ }\href
  {https://doi.org/10.1103/PhysRevB.103.214416} {\bibfield  {journal} {\bibinfo
   {journal} {Phys. Rev. B}\ }\textbf {\bibinfo {volume} {103}},\ \bibinfo
  {pages} {214416} (\bibinfo {year} {2021})}\BibitemShut {NoStop}%
\bibitem [{\citenamefont {Singh}\ \emph {et~al.}(2023)\citenamefont {Singh},
  \citenamefont {Stavropoulos},\ and\ \citenamefont {Perkins}}]{sin23}%
  \BibitemOpen
  \bibfield  {author} {\bibinfo {author} {\bibfnamefont {S.}~\bibnamefont
  {Singh}}, \bibinfo {author} {\bibfnamefont {P.~P.}\ \bibnamefont
  {Stavropoulos}},\ and\ \bibinfo {author} {\bibfnamefont {N.~B.}\ \bibnamefont
  {Perkins}},\ }\bibfield  {title} {\bibinfo {title} {Phonon dynamics in the
  generalized kitaev spin liquid},\ }\href
  {https://doi.org/10.1103/PhysRevB.107.214428} {\bibfield  {journal} {\bibinfo
   {journal} {Phys. Rev. B}\ }\textbf {\bibinfo {volume} {107}},\ \bibinfo
  {pages} {214428} (\bibinfo {year} {2023})}\BibitemShut {NoStop}%
\bibitem [{\citenamefont {Singh}\ \emph {et~al.}(2024)\citenamefont {Singh},
  \citenamefont {Stavropoulos},\ and\ \citenamefont {Perkins}}]{Singh2024}%
  \BibitemOpen
  \bibfield  {author} {\bibinfo {author} {\bibfnamefont {S.}~\bibnamefont
  {Singh}}, \bibinfo {author} {\bibfnamefont {P.~P.}\ \bibnamefont
  {Stavropoulos}},\ and\ \bibinfo {author} {\bibfnamefont {N.~B.}\ \bibnamefont
  {Perkins}},\ }\bibfield  {title} {\bibinfo {title} {Phonon dynamics in the
  chiral kitaev spin liquid},\ }\href
  {https://doi.org/10.1103/PhysRevB.110.094431} {\bibfield  {journal} {\bibinfo
   {journal} {Phys. Rev. B}\ }\textbf {\bibinfo {volume} {110}},\ \bibinfo
  {pages} {094431} (\bibinfo {year} {2024})}\BibitemShut {NoStop}%
\bibitem [{\citenamefont {Dantas}\ \emph {et~al.}(2024)\citenamefont {Dantas},
  \citenamefont {Kao},\ and\ \citenamefont {Perkins}}]{Dantas2024}%
  \BibitemOpen
  \bibfield  {author} {\bibinfo {author} {\bibfnamefont {V.}~\bibnamefont
  {Dantas}}, \bibinfo {author} {\bibfnamefont {W.-H.}\ \bibnamefont {Kao}},\
  and\ \bibinfo {author} {\bibfnamefont {N.~B.}\ \bibnamefont {Perkins}},\
  }\bibfield  {title} {\bibinfo {title} {Phonon dynamics in the site-disordered
  kitaev spin liquid},\ }\href {https://doi.org/10.1103/PhysRevB.110.104425}
  {\bibfield  {journal} {\bibinfo  {journal} {Phys. Rev. B}\ }\textbf {\bibinfo
  {volume} {110}},\ \bibinfo {pages} {104425} (\bibinfo {year}
  {2024})}\BibitemShut {NoStop}%
\bibitem [{\citenamefont {Cui}\ \emph {et~al.}(2017)\citenamefont {Cui},
  \citenamefont {Zheng}, \citenamefont {Ran}, \citenamefont {Wen},
  \citenamefont {Liu}, \citenamefont {Liu}, \citenamefont {Guo},\ and\
  \citenamefont {Yu}}]{cui17}%
  \BibitemOpen
  \bibfield  {author} {\bibinfo {author} {\bibfnamefont {Y.}~\bibnamefont
  {Cui}}, \bibinfo {author} {\bibfnamefont {J.}~\bibnamefont {Zheng}}, \bibinfo
  {author} {\bibfnamefont {K.}~\bibnamefont {Ran}}, \bibinfo {author}
  {\bibfnamefont {J.}~\bibnamefont {Wen}}, \bibinfo {author} {\bibfnamefont
  {Z.-X.}\ \bibnamefont {Liu}}, \bibinfo {author} {\bibfnamefont
  {B.}~\bibnamefont {Liu}}, \bibinfo {author} {\bibfnamefont {W.}~\bibnamefont
  {Guo}},\ and\ \bibinfo {author} {\bibfnamefont {W.}~\bibnamefont {Yu}},\
  }\bibfield  {title} {\bibinfo {title} {High-pressure magnetization and nmr
  studies of $\ensuremath{\alpha}\text{\ensuremath{-}}{\mathrm{rucl}}_{3}$},\
  }\href {https://doi.org/10.1103/PhysRevB.96.205147} {\bibfield  {journal}
  {\bibinfo  {journal} {Phys. Rev. B}\ }\textbf {\bibinfo {volume} {96}},\
  \bibinfo {pages} {205147} (\bibinfo {year} {2017})}\BibitemShut {NoStop}%
\bibitem [{\citenamefont {Biesner}\ \emph {et~al.}(2018)\citenamefont
  {Biesner}, \citenamefont {Biswas}, \citenamefont {Li}, \citenamefont {Saito},
  \citenamefont {Pustogow}, \citenamefont {Altmeyer}, \citenamefont {Wolter},
  \citenamefont {B\"uchner}, \citenamefont {Roslova}, \citenamefont {Doert},
  \citenamefont {Winter}, \citenamefont {Valent\'{\i}},\ and\ \citenamefont
  {Dressel}}]{bie18}%
  \BibitemOpen
  \bibfield  {author} {\bibinfo {author} {\bibfnamefont {T.}~\bibnamefont
  {Biesner}}, \bibinfo {author} {\bibfnamefont {S.}~\bibnamefont {Biswas}},
  \bibinfo {author} {\bibfnamefont {W.}~\bibnamefont {Li}}, \bibinfo {author}
  {\bibfnamefont {Y.}~\bibnamefont {Saito}}, \bibinfo {author} {\bibfnamefont
  {A.}~\bibnamefont {Pustogow}}, \bibinfo {author} {\bibfnamefont
  {M.}~\bibnamefont {Altmeyer}}, \bibinfo {author} {\bibfnamefont {A.~U.~B.}\
  \bibnamefont {Wolter}}, \bibinfo {author} {\bibfnamefont {B.}~\bibnamefont
  {B\"uchner}}, \bibinfo {author} {\bibfnamefont {M.}~\bibnamefont {Roslova}},
  \bibinfo {author} {\bibfnamefont {T.}~\bibnamefont {Doert}}, \bibinfo
  {author} {\bibfnamefont {S.~M.}\ \bibnamefont {Winter}}, \bibinfo {author}
  {\bibfnamefont {R.}~\bibnamefont {Valent\'{\i}}},\ and\ \bibinfo {author}
  {\bibfnamefont {M.}~\bibnamefont {Dressel}},\ }\bibfield  {title} {\bibinfo
  {title} {Detuning the honeycomb of
  $\ensuremath{\alpha}\text{\ensuremath{-}}{\mathrm{rucl}}_{3}$:
  Pressure-dependent optical studies reveal broken symmetry},\ }\href
  {https://doi.org/10.1103/PhysRevB.97.220401} {\bibfield  {journal} {\bibinfo
  {journal} {Phys. Rev. B}\ }\textbf {\bibinfo {volume} {97}},\ \bibinfo
  {pages} {220401} (\bibinfo {year} {2018})}\BibitemShut {NoStop}%
\bibitem [{\citenamefont {Bastien}\ \emph {et~al.}(2018)\citenamefont
  {Bastien}, \citenamefont {Garbarino}, \citenamefont {Yadav}, \citenamefont
  {Martinez-Casado}, \citenamefont {Beltr\'an~Rodr\'{\i}guez}, \citenamefont
  {Stahl}, \citenamefont {Kusch}, \citenamefont {Limandri}, \citenamefont
  {Ray}, \citenamefont {Lampen-Kelley}, \citenamefont {Mandrus}, \citenamefont
  {Nagler}, \citenamefont {Roslova}, \citenamefont {Isaeva}, \citenamefont
  {Doert}, \citenamefont {Hozoi}, \citenamefont {Wolter}, \citenamefont
  {B\"uchner}, \citenamefont {Geck},\ and\ \citenamefont {van~den
  Brink}}]{bas18}%
  \BibitemOpen
  \bibfield  {author} {\bibinfo {author} {\bibfnamefont {G.}~\bibnamefont
  {Bastien}}, \bibinfo {author} {\bibfnamefont {G.}~\bibnamefont {Garbarino}},
  \bibinfo {author} {\bibfnamefont {R.}~\bibnamefont {Yadav}}, \bibinfo
  {author} {\bibfnamefont {F.~J.}\ \bibnamefont {Martinez-Casado}}, \bibinfo
  {author} {\bibfnamefont {R.}~\bibnamefont {Beltr\'an~Rodr\'{\i}guez}},
  \bibinfo {author} {\bibfnamefont {Q.}~\bibnamefont {Stahl}}, \bibinfo
  {author} {\bibfnamefont {M.}~\bibnamefont {Kusch}}, \bibinfo {author}
  {\bibfnamefont {S.~P.}\ \bibnamefont {Limandri}}, \bibinfo {author}
  {\bibfnamefont {R.}~\bibnamefont {Ray}}, \bibinfo {author} {\bibfnamefont
  {P.}~\bibnamefont {Lampen-Kelley}}, \bibinfo {author} {\bibfnamefont {D.~G.}\
  \bibnamefont {Mandrus}}, \bibinfo {author} {\bibfnamefont {S.~E.}\
  \bibnamefont {Nagler}}, \bibinfo {author} {\bibfnamefont {M.}~\bibnamefont
  {Roslova}}, \bibinfo {author} {\bibfnamefont {A.}~\bibnamefont {Isaeva}},
  \bibinfo {author} {\bibfnamefont {T.}~\bibnamefont {Doert}}, \bibinfo
  {author} {\bibfnamefont {L.}~\bibnamefont {Hozoi}}, \bibinfo {author}
  {\bibfnamefont {A.~U.~B.}\ \bibnamefont {Wolter}}, \bibinfo {author}
  {\bibfnamefont {B.}~\bibnamefont {B\"uchner}}, \bibinfo {author}
  {\bibfnamefont {J.}~\bibnamefont {Geck}},\ and\ \bibinfo {author}
  {\bibfnamefont {J.}~\bibnamefont {van~den Brink}},\ }\bibfield  {title}
  {\bibinfo {title} {Pressure-induced dimerization and valence bond crystal
  formation in the kitaev-heisenberg magnet
  $\ensuremath{\alpha}\text{\ensuremath{-}}{\mathrm{rucl}}_{3}$},\ }\href
  {https://doi.org/10.1103/PhysRevB.97.241108} {\bibfield  {journal} {\bibinfo
  {journal} {Phys. Rev. B}\ }\textbf {\bibinfo {volume} {97}},\ \bibinfo
  {pages} {241108} (\bibinfo {year} {2018})}\BibitemShut {NoStop}%
\bibitem [{\citenamefont {Wang}\ \emph {et~al.}(2023)\citenamefont {Wang},
  \citenamefont {Zhu}, \citenamefont {Qureshi}, \citenamefont {Beauvois},
  \citenamefont {Song}, \citenamefont {Mueller}, \citenamefont {Brückel},\
  and\ \citenamefont {Su}}]{wan23}%
  \BibitemOpen
  \bibfield  {author} {\bibinfo {author} {\bibfnamefont {X.}~\bibnamefont
  {Wang}}, \bibinfo {author} {\bibfnamefont {F.}~\bibnamefont {Zhu}}, \bibinfo
  {author} {\bibfnamefont {N.}~\bibnamefont {Qureshi}}, \bibinfo {author}
  {\bibfnamefont {K.}~\bibnamefont {Beauvois}}, \bibinfo {author}
  {\bibfnamefont {J.}~\bibnamefont {Song}}, \bibinfo {author} {\bibfnamefont
  {T.}~\bibnamefont {Mueller}}, \bibinfo {author} {\bibfnamefont
  {T.}~\bibnamefont {Brückel}},\ and\ \bibinfo {author} {\bibfnamefont
  {Y.}~\bibnamefont {Su}},\ }\bibfield  {title} {\bibinfo {title} {Hydrostatic
  pressure effects in the kitaev quantum magnet
  $\ensuremath{\alpha}\ensuremath{-}{\mathrm{rucl}}_{3}$: A single-crystal
  neutron diffraction study},\ }\href@noop {} {\bibfield  {journal} {\bibinfo
  {journal} {arXiv:2304.00632}\ } (\bibinfo {year} {2023})}\BibitemShut
  {NoStop}%
\bibitem [{\citenamefont {Wolf}\ \emph {et~al.}(2022)\citenamefont {Wolf},
  \citenamefont {Kaib}, \citenamefont {Razpopov}, \citenamefont {Biswas},
  \citenamefont {Riedl}, \citenamefont {Winter}, \citenamefont {Valent\'{\i}},
  \citenamefont {Saito}, \citenamefont {Hartmann}, \citenamefont {Vinokurova},
  \citenamefont {Doert}, \citenamefont {Isaeva}, \citenamefont {Bastien},
  \citenamefont {Wolter}, \citenamefont {B\"uchner},\ and\ \citenamefont
  {Lang}}]{wolf22}%
  \BibitemOpen
  \bibfield  {author} {\bibinfo {author} {\bibfnamefont {B.}~\bibnamefont
  {Wolf}}, \bibinfo {author} {\bibfnamefont {D.~A.~S.}\ \bibnamefont {Kaib}},
  \bibinfo {author} {\bibfnamefont {A.}~\bibnamefont {Razpopov}}, \bibinfo
  {author} {\bibfnamefont {S.}~\bibnamefont {Biswas}}, \bibinfo {author}
  {\bibfnamefont {K.}~\bibnamefont {Riedl}}, \bibinfo {author} {\bibfnamefont
  {S.~M.}\ \bibnamefont {Winter}}, \bibinfo {author} {\bibfnamefont
  {R.}~\bibnamefont {Valent\'{\i}}}, \bibinfo {author} {\bibfnamefont
  {Y.}~\bibnamefont {Saito}}, \bibinfo {author} {\bibfnamefont
  {S.}~\bibnamefont {Hartmann}}, \bibinfo {author} {\bibfnamefont
  {E.}~\bibnamefont {Vinokurova}}, \bibinfo {author} {\bibfnamefont
  {T.}~\bibnamefont {Doert}}, \bibinfo {author} {\bibfnamefont
  {A.}~\bibnamefont {Isaeva}}, \bibinfo {author} {\bibfnamefont
  {G.}~\bibnamefont {Bastien}}, \bibinfo {author} {\bibfnamefont {A.~U.~B.}\
  \bibnamefont {Wolter}}, \bibinfo {author} {\bibfnamefont {B.}~\bibnamefont
  {B\"uchner}},\ and\ \bibinfo {author} {\bibfnamefont {M.}~\bibnamefont
  {Lang}},\ }\bibfield  {title} {\bibinfo {title} {Combined experimental and
  theoretical study of hydrostatic he-gas pressure effects in
  $\ensuremath{\alpha}\text{\ensuremath{-}}{\mathrm{rucl}}_{3}$},\ }\href
  {https://doi.org/10.1103/PhysRevB.106.134432} {\bibfield  {journal} {\bibinfo
   {journal} {Phys. Rev. B}\ }\textbf {\bibinfo {volume} {106}},\ \bibinfo
  {pages} {134432} (\bibinfo {year} {2022})}\BibitemShut {NoStop}%
\bibitem [{\citenamefont {Stahl}\ \emph {et~al.}(2024)\citenamefont {Stahl},
  \citenamefont {Ritschel}, \citenamefont {Gararino}, \citenamefont {Cova},
  \citenamefont {Isaeva}, \citenamefont {Doert},\ and\ \citenamefont
  {Geck}}]{stanco24}%
  \BibitemOpen
  \bibfield  {author} {\bibinfo {author} {\bibfnamefont {Q.}~\bibnamefont
  {Stahl}}, \bibinfo {author} {\bibfnamefont {T.}~\bibnamefont {Ritschel}},
  \bibinfo {author} {\bibfnamefont {G.}~\bibnamefont {Gararino}}, \bibinfo
  {author} {\bibfnamefont {F.}~\bibnamefont {Cova}}, \bibinfo {author}
  {\bibfnamefont {A.}~\bibnamefont {Isaeva}}, \bibinfo {author} {\bibfnamefont
  {T.}~\bibnamefont {Doert}},\ and\ \bibinfo {author} {\bibfnamefont
  {J.}~\bibnamefont {Geck}},\ }\bibfield  {title} {\bibinfo {title}
  {Pressure-tuning of $\alpha$-rucl$_3$ towards a quantum spin liquid},\
  }\href@noop {} {\bibfield  {journal} {\bibinfo  {journal} {Nat. Commun.}\
  }\textbf {\bibinfo {volume} {15}},\ \bibinfo {pages} {8142} (\bibinfo {year}
  {2024})}\BibitemShut {NoStop}%
\bibitem [{\citenamefont {Daniels}\ and\ \citenamefont
  {Ryschkewitsch}(1983)}]{danielssimple1983}%
  \BibitemOpen
  \bibfield  {author} {\bibinfo {author} {\bibfnamefont {W.~B.}\ \bibnamefont
  {Daniels}}\ and\ \bibinfo {author} {\bibfnamefont {M.~G.}\ \bibnamefont
  {Ryschkewitsch}},\ }\bibfield  {title} {\bibinfo {title} {Simple double
  diaphragm press for diamond anvil cells at low temperatures},\ }\href
  {https://doi.org/10.1063/1.1137223} {\bibfield  {journal} {\bibinfo
  {journal} {Rev. Sci. Instrum.}\ }\textbf {\bibinfo {volume} {54}},\ \bibinfo
  {pages} {115} (\bibinfo {year} {1983})}\BibitemShut {NoStop}%
\bibitem [{\citenamefont {Wang}\ \emph {et~al.}(2018)\citenamefont {Wang},
  \citenamefont {Guo}, \citenamefont {Tafti}, \citenamefont {Hegg},
  \citenamefont {Sen}, \citenamefont {Sidorov}, \citenamefont {Wang},
  \citenamefont {Cai}, \citenamefont {Yi}, \citenamefont {Zhou}, \citenamefont
  {Wang}, \citenamefont {Zhang}, \citenamefont {Yang}, \citenamefont {Li},
  \citenamefont {Li}, \citenamefont {Li}, \citenamefont {Liu}, \citenamefont
  {Shi}, \citenamefont {Ku}, \citenamefont {Wu}, \citenamefont {Cava},\ and\
  \citenamefont {Sun}}]{wangprb18}%
  \BibitemOpen
  \bibfield  {author} {\bibinfo {author} {\bibfnamefont {Z.}~\bibnamefont
  {Wang}}, \bibinfo {author} {\bibfnamefont {J.}~\bibnamefont {Guo}}, \bibinfo
  {author} {\bibfnamefont {F.~F.}\ \bibnamefont {Tafti}}, \bibinfo {author}
  {\bibfnamefont {A.}~\bibnamefont {Hegg}}, \bibinfo {author} {\bibfnamefont
  {S.}~\bibnamefont {Sen}}, \bibinfo {author} {\bibfnamefont {V.~A.}\
  \bibnamefont {Sidorov}}, \bibinfo {author} {\bibfnamefont {L.}~\bibnamefont
  {Wang}}, \bibinfo {author} {\bibfnamefont {S.}~\bibnamefont {Cai}}, \bibinfo
  {author} {\bibfnamefont {W.}~\bibnamefont {Yi}}, \bibinfo {author}
  {\bibfnamefont {Y.}~\bibnamefont {Zhou}}, \bibinfo {author} {\bibfnamefont
  {H.}~\bibnamefont {Wang}}, \bibinfo {author} {\bibfnamefont {S.}~\bibnamefont
  {Zhang}}, \bibinfo {author} {\bibfnamefont {K.}~\bibnamefont {Yang}},
  \bibinfo {author} {\bibfnamefont {A.}~\bibnamefont {Li}}, \bibinfo {author}
  {\bibfnamefont {X.}~\bibnamefont {Li}}, \bibinfo {author} {\bibfnamefont
  {Y.}~\bibnamefont {Li}}, \bibinfo {author} {\bibfnamefont {J.}~\bibnamefont
  {Liu}}, \bibinfo {author} {\bibfnamefont {Y.}~\bibnamefont {Shi}}, \bibinfo
  {author} {\bibfnamefont {W.}~\bibnamefont {Ku}}, \bibinfo {author}
  {\bibfnamefont {Q.}~\bibnamefont {Wu}}, \bibinfo {author} {\bibfnamefont
  {R.~J.}\ \bibnamefont {Cava}},\ and\ \bibinfo {author} {\bibfnamefont
  {L.}~\bibnamefont {Sun}},\ }\bibfield  {title} {\bibinfo {title}
  {Pressure-induced melting of magnetic order and emergence of a new quantum
  state in
  $\ensuremath{\alpha}\text{\ensuremath{-}}\mathrm{RuC}{\mathrm{l}}_{3}$},\
  }\href {https://doi.org/10.1103/PhysRevB.97.245149} {\bibfield  {journal}
  {\bibinfo  {journal} {Phys. Rev. B}\ }\textbf {\bibinfo {volume} {97}},\
  \bibinfo {pages} {245149} (\bibinfo {year} {2018})}\BibitemShut {NoStop}%
\bibitem [{\citenamefont {Momma}\ and\ \citenamefont
  {Izumi}(2011)}]{mommavesta32011}%
  \BibitemOpen
  \bibfield  {author} {\bibinfo {author} {\bibfnamefont {K.}~\bibnamefont
  {Momma}}\ and\ \bibinfo {author} {\bibfnamefont {F.}~\bibnamefont {Izumi}},\
  }\bibfield  {title} {\bibinfo {title} {{VESTA3} for three-dimensional
  visualization of crystal, volumetric and morphology data},\ }\href
  {https://doi.org/10.1107/S0021889811038970} {\bibfield  {journal} {\bibinfo
  {journal} {J. Appl. Crystallogr.}\ }\textbf {\bibinfo {volume} {44}},\
  \bibinfo {pages} {1272} (\bibinfo {year} {2011})}\BibitemShut {NoStop}%
\bibitem [{\citenamefont {Reschke}\ \emph {et~al.}(2018)\citenamefont
  {Reschke}, \citenamefont {Mayr}, \citenamefont {Widmann}, \citenamefont {von
  Nidda}, \citenamefont {Tsurkan}, \citenamefont {Eremin}, \citenamefont {Do},
  \citenamefont {Choi}, \citenamefont {Wang},\ and\ \citenamefont
  {Loidl}}]{reschkesubgap2018}%
  \BibitemOpen
  \bibfield  {author} {\bibinfo {author} {\bibfnamefont {S.}~\bibnamefont
  {Reschke}}, \bibinfo {author} {\bibfnamefont {F.}~\bibnamefont {Mayr}},
  \bibinfo {author} {\bibfnamefont {S.}~\bibnamefont {Widmann}}, \bibinfo
  {author} {\bibfnamefont {H.-A.~K.}\ \bibnamefont {von Nidda}}, \bibinfo
  {author} {\bibfnamefont {V.}~\bibnamefont {Tsurkan}}, \bibinfo {author}
  {\bibfnamefont {M.~V.}\ \bibnamefont {Eremin}}, \bibinfo {author}
  {\bibfnamefont {S.-H.}\ \bibnamefont {Do}}, \bibinfo {author} {\bibfnamefont
  {K.-Y.}\ \bibnamefont {Choi}}, \bibinfo {author} {\bibfnamefont
  {Z.}~\bibnamefont {Wang}},\ and\ \bibinfo {author} {\bibfnamefont
  {A.}~\bibnamefont {Loidl}},\ }\bibfield  {title} {\bibinfo {title} {Sub-gap
  optical response in the {Kitaev} spin-liquid candidate \textit{$\alpha$}
  -{RuCl} $_{\textrm{3}}$},\ }\href {https://doi.org/10.1088/1361-648X/aae805}
  {\bibfield  {journal} {\bibinfo  {journal} {J. Phys.: Condens. Matter}\
  }\textbf {\bibinfo {volume} {30}},\ \bibinfo {pages} {475604} (\bibinfo
  {year} {2018})}\BibitemShut {NoStop}%
\bibitem [{\citenamefont {Mi}\ \emph {et~al.}(2021)\citenamefont {Mi},
  \citenamefont {Wang}, \citenamefont {Gui}, \citenamefont {Pi}, \citenamefont
  {Zheng}, \citenamefont {Yang}, \citenamefont {Gan}, \citenamefont {Wang},
  \citenamefont {Li}, \citenamefont {Wang}, \citenamefont {Zhang},
  \citenamefont {Su}, \citenamefont {Chai},\ and\ \citenamefont {He}}]{mi21}%
  \BibitemOpen
  \bibfield  {author} {\bibinfo {author} {\bibfnamefont {X.}~\bibnamefont
  {Mi}}, \bibinfo {author} {\bibfnamefont {X.}~\bibnamefont {Wang}}, \bibinfo
  {author} {\bibfnamefont {H.}~\bibnamefont {Gui}}, \bibinfo {author}
  {\bibfnamefont {M.}~\bibnamefont {Pi}}, \bibinfo {author} {\bibfnamefont
  {T.}~\bibnamefont {Zheng}}, \bibinfo {author} {\bibfnamefont
  {K.}~\bibnamefont {Yang}}, \bibinfo {author} {\bibfnamefont {Y.}~\bibnamefont
  {Gan}}, \bibinfo {author} {\bibfnamefont {P.}~\bibnamefont {Wang}}, \bibinfo
  {author} {\bibfnamefont {A.}~\bibnamefont {Li}}, \bibinfo {author}
  {\bibfnamefont {A.}~\bibnamefont {Wang}}, \bibinfo {author} {\bibfnamefont
  {L.}~\bibnamefont {Zhang}}, \bibinfo {author} {\bibfnamefont
  {Y.}~\bibnamefont {Su}}, \bibinfo {author} {\bibfnamefont {Y.}~\bibnamefont
  {Chai}},\ and\ \bibinfo {author} {\bibfnamefont {M.}~\bibnamefont {He}},\
  }\bibfield  {title} {\bibinfo {title} {Stacking faults in
  $\ensuremath{\alpha}\text{\ensuremath{-}}{\mathrm{rucl}}_{3}$ revealed by
  local electric polarization},\ }\href
  {https://doi.org/10.1103/PhysRevB.103.174413} {\bibfield  {journal} {\bibinfo
   {journal} {Phys. Rev. B}\ }\textbf {\bibinfo {volume} {103}},\ \bibinfo
  {pages} {174413} (\bibinfo {year} {2021})}\BibitemShut {NoStop}%
\bibitem [{\citenamefont {Cao}\ \emph {et~al.}(2016)\citenamefont {Cao},
  \citenamefont {Banerjee}, \citenamefont {Yan}, \citenamefont {Bridges},
  \citenamefont {Lumsden}, \citenamefont {Mandrus}, \citenamefont {Tennant},
  \citenamefont {Chakoumakos},\ and\ \citenamefont
  {Nagler}}]{caolowtemperature2016}%
  \BibitemOpen
  \bibfield  {author} {\bibinfo {author} {\bibfnamefont {H.~B.}\ \bibnamefont
  {Cao}}, \bibinfo {author} {\bibfnamefont {A.}~\bibnamefont {Banerjee}},
  \bibinfo {author} {\bibfnamefont {J.-Q.}\ \bibnamefont {Yan}}, \bibinfo
  {author} {\bibfnamefont {C.~A.}\ \bibnamefont {Bridges}}, \bibinfo {author}
  {\bibfnamefont {M.~D.}\ \bibnamefont {Lumsden}}, \bibinfo {author}
  {\bibfnamefont {D.~G.}\ \bibnamefont {Mandrus}}, \bibinfo {author}
  {\bibfnamefont {D.~A.}\ \bibnamefont {Tennant}}, \bibinfo {author}
  {\bibfnamefont {B.~C.}\ \bibnamefont {Chakoumakos}},\ and\ \bibinfo {author}
  {\bibfnamefont {S.~E.}\ \bibnamefont {Nagler}},\ }\bibfield  {title}
  {\bibinfo {title} {Low-temperature crystal and magnetic structure of
  $\ensuremath{\alpha}\ensuremath{-}{\mathrm{rucl}}_{3}$},\ }\href
  {https://doi.org/10.1103/PhysRevB.93.134423} {\bibfield  {journal} {\bibinfo
  {journal} {Phys. Rev. B}\ }\textbf {\bibinfo {volume} {93}},\ \bibinfo
  {pages} {134423} (\bibinfo {year} {2016})}\BibitemShut {NoStop}%
\bibitem [{\citenamefont {Mu}\ \emph {et~al.}(2022)\citenamefont {Mu},
  \citenamefont {Dixit}, \citenamefont {Wang}, \citenamefont {Abernathy},
  \citenamefont {Cao}, \citenamefont {Nagler}, \citenamefont {Yan},
  \citenamefont {Lampen-Kelley}, \citenamefont {Mandrus}, \citenamefont
  {Polanco}, \citenamefont {Liang}, \citenamefont {Halász}, \citenamefont
  {Cheng}, \citenamefont {Banerjee},\ and\ \citenamefont
  {Berlijn}}]{murole2022}%
  \BibitemOpen
  \bibfield  {author} {\bibinfo {author} {\bibfnamefont {S.}~\bibnamefont
  {Mu}}, \bibinfo {author} {\bibfnamefont {K.~D.}\ \bibnamefont {Dixit}},
  \bibinfo {author} {\bibfnamefont {X.}~\bibnamefont {Wang}}, \bibinfo {author}
  {\bibfnamefont {D.~L.}\ \bibnamefont {Abernathy}}, \bibinfo {author}
  {\bibfnamefont {H.}~\bibnamefont {Cao}}, \bibinfo {author} {\bibfnamefont
  {S.~E.}\ \bibnamefont {Nagler}}, \bibinfo {author} {\bibfnamefont
  {J.}~\bibnamefont {Yan}}, \bibinfo {author} {\bibfnamefont {P.}~\bibnamefont
  {Lampen-Kelley}}, \bibinfo {author} {\bibfnamefont {D.}~\bibnamefont
  {Mandrus}}, \bibinfo {author} {\bibfnamefont {C.~A.}\ \bibnamefont
  {Polanco}}, \bibinfo {author} {\bibfnamefont {L.}~\bibnamefont {Liang}},
  \bibinfo {author} {\bibfnamefont {G.~B.}\ \bibnamefont {Halász}}, \bibinfo
  {author} {\bibfnamefont {Y.}~\bibnamefont {Cheng}}, \bibinfo {author}
  {\bibfnamefont {A.}~\bibnamefont {Banerjee}},\ and\ \bibinfo {author}
  {\bibfnamefont {T.}~\bibnamefont {Berlijn}},\ }\bibfield  {title} {\bibinfo
  {title} {Role of the third dimension in searching for {Majorana} fermions in
  $\alpha$ - {RuCl} 3 via phonons},\ }\href
  {https://doi.org/10.1103/PhysRevResearch.4.013067} {\bibfield  {journal}
  {\bibinfo  {journal} {Phys. Rev. Research}\ }\textbf {\bibinfo {volume}
  {4}},\ \bibinfo {pages} {013067} (\bibinfo {year} {2022})}\BibitemShut
  {NoStop}%
\bibitem [{\citenamefont {Lebert}\ \emph {et~al.}(2022)\citenamefont {Lebert},
  \citenamefont {Kim}, \citenamefont {Prishchenko}, \citenamefont {Tsirlin},
  \citenamefont {Said}, \citenamefont {Alatas},\ and\ \citenamefont
  {Kim}}]{lebertacoustic2022}%
  \BibitemOpen
  \bibfield  {author} {\bibinfo {author} {\bibfnamefont {B.~W.}\ \bibnamefont
  {Lebert}}, \bibinfo {author} {\bibfnamefont {S.}~\bibnamefont {Kim}},
  \bibinfo {author} {\bibfnamefont {D.~A.}\ \bibnamefont {Prishchenko}},
  \bibinfo {author} {\bibfnamefont {A.~A.}\ \bibnamefont {Tsirlin}}, \bibinfo
  {author} {\bibfnamefont {A.~H.}\ \bibnamefont {Said}}, \bibinfo {author}
  {\bibfnamefont {A.}~\bibnamefont {Alatas}},\ and\ \bibinfo {author}
  {\bibfnamefont {Y.-J.}\ \bibnamefont {Kim}},\ }\bibfield  {title} {\bibinfo
  {title} {Acoustic phonon dispersion of $\alpha$- {RuCl} 3},\ }\href
  {https://doi.org/10.1103/PhysRevB.106.L041102} {\bibfield  {journal}
  {\bibinfo  {journal} {Phys. Rev. B}\ }\textbf {\bibinfo {volume} {106}},\
  \bibinfo {pages} {L041102} (\bibinfo {year} {2022})}\BibitemShut {NoStop}%
\bibitem [{\citenamefont {Yamauchi}\ \emph {et~al.}(2018)\citenamefont
  {Yamauchi}, \citenamefont {Hiraishi}, \citenamefont {Okabe}, \citenamefont
  {Takeshita}, \citenamefont {Koda}, \citenamefont {Kojima}, \citenamefont
  {Kadono},\ and\ \citenamefont {Tanaka}}]{yamauchilocal2018}%
  \BibitemOpen
  \bibfield  {author} {\bibinfo {author} {\bibfnamefont {I.}~\bibnamefont
  {Yamauchi}}, \bibinfo {author} {\bibfnamefont {M.}~\bibnamefont {Hiraishi}},
  \bibinfo {author} {\bibfnamefont {H.}~\bibnamefont {Okabe}}, \bibinfo
  {author} {\bibfnamefont {S.}~\bibnamefont {Takeshita}}, \bibinfo {author}
  {\bibfnamefont {A.}~\bibnamefont {Koda}}, \bibinfo {author} {\bibfnamefont
  {K.~M.}\ \bibnamefont {Kojima}}, \bibinfo {author} {\bibfnamefont
  {R.}~\bibnamefont {Kadono}},\ and\ \bibinfo {author} {\bibfnamefont
  {H.}~\bibnamefont {Tanaka}},\ }\bibfield  {title} {\bibinfo {title} {Local
  spin structure of the $\ensuremath{\alpha}\ensuremath{-}{\mathrm{rucl}}_{3}$
  honeycomb-lattice magnet observed via muon spin rotation/relaxation},\ }\href
  {https://doi.org/10.1103/PhysRevB.97.134410} {\bibfield  {journal} {\bibinfo
  {journal} {Phys. Rev. B}\ }\textbf {\bibinfo {volume} {97}},\ \bibinfo
  {pages} {134410} (\bibinfo {year} {2018})}\BibitemShut {NoStop}%
\bibitem [{\citenamefont {Wolf}\ \emph {et~al.}(2001)\citenamefont {Wolf},
  \citenamefont {Lüthi}, \citenamefont {Schmidt}, \citenamefont {Schwenk},
  \citenamefont {Sieling}, \citenamefont {Zherlitsyn},\ and\ \citenamefont
  {Kouroudis}}]{wolfnew2001}%
  \BibitemOpen
  \bibfield  {author} {\bibinfo {author} {\bibfnamefont {B.}~\bibnamefont
  {Wolf}}, \bibinfo {author} {\bibfnamefont {B.}~\bibnamefont {Lüthi}},
  \bibinfo {author} {\bibfnamefont {S.}~\bibnamefont {Schmidt}}, \bibinfo
  {author} {\bibfnamefont {H.}~\bibnamefont {Schwenk}}, \bibinfo {author}
  {\bibfnamefont {M.}~\bibnamefont {Sieling}}, \bibinfo {author} {\bibfnamefont
  {S.}~\bibnamefont {Zherlitsyn}},\ and\ \bibinfo {author} {\bibfnamefont
  {I.}~\bibnamefont {Kouroudis}},\ }\bibfield  {title} {\bibinfo {title} {New
  experimental techniques for pulsed magnetic fields – {ESR} and
  ultrasonics},\ }\href
  {https://doi.org/https://doi.org/10.1016/S0921-4526(00)00729-8} {\bibfield
  {journal} {\bibinfo  {journal} {Physica B: Condensed Matter}\ }\textbf
  {\bibinfo {volume} {294-295}},\ \bibinfo {pages} {612} (\bibinfo {year}
  {2001})}\BibitemShut {NoStop}%
\bibitem [{\citenamefont {Staško}\ \emph {et~al.}(2020)\citenamefont
  {Staško}, \citenamefont {Prchal}, \citenamefont {Klicpera}, \citenamefont
  {Aoki},\ and\ \citenamefont {Murata}}]{staskopressure2020}%
  \BibitemOpen
  \bibfield  {author} {\bibinfo {author} {\bibfnamefont {D.}~\bibnamefont
  {Staško}}, \bibinfo {author} {\bibfnamefont {J.}~\bibnamefont {Prchal}},
  \bibinfo {author} {\bibfnamefont {M.}~\bibnamefont {Klicpera}}, \bibinfo
  {author} {\bibfnamefont {S.}~\bibnamefont {Aoki}},\ and\ \bibinfo {author}
  {\bibfnamefont {K.}~\bibnamefont {Murata}},\ }\bibfield  {title} {\bibinfo
  {title} {Pressure media for high pressure experiments, {Daphne} {Oil} 7000
  series},\ }\href {https://doi.org/10.1080/08957959.2020.1825706} {\bibfield
  {journal} {\bibinfo  {journal} {High Pressure Research}\ }\textbf {\bibinfo
  {volume} {40}},\ \bibinfo {pages} {525} (\bibinfo {year} {2020})}\BibitemShut
  {NoStop}%
\bibitem [{\citenamefont {Mombetsu}\ \emph {et~al.}(2016)\citenamefont
  {Mombetsu}, \citenamefont {Murazumi}, \citenamefont {Hidaka}, \citenamefont
  {Yanagisawa}, \citenamefont {Amitsuka}, \citenamefont {Ho},\ and\
  \citenamefont {Maple}}]{mombetsustudy2016}%
  \BibitemOpen
  \bibfield  {author} {\bibinfo {author} {\bibfnamefont {S.}~\bibnamefont
  {Mombetsu}}, \bibinfo {author} {\bibfnamefont {T.}~\bibnamefont {Murazumi}},
  \bibinfo {author} {\bibfnamefont {H.}~\bibnamefont {Hidaka}}, \bibinfo
  {author} {\bibfnamefont {T.}~\bibnamefont {Yanagisawa}}, \bibinfo {author}
  {\bibfnamefont {H.}~\bibnamefont {Amitsuka}}, \bibinfo {author}
  {\bibfnamefont {P.-C.}\ \bibnamefont {Ho}},\ and\ \bibinfo {author}
  {\bibfnamefont {M.~B.}\ \bibnamefont {Maple}},\ }\bibfield  {title} {\bibinfo
  {title} {Study of localized character of 4 f electrons and ultrasonic
  dispersions in {SmOs} 4 {Sb} 12 by high-pressure high-frequency ultrasonic
  measurements},\ }\href {https://doi.org/10.1103/PhysRevB.94.085142}
  {\bibfield  {journal} {\bibinfo  {journal} {Phys. Rev. B}\ }\textbf {\bibinfo
  {volume} {94}},\ \bibinfo {pages} {085142} (\bibinfo {year}
  {2016})}\BibitemShut {NoStop}%
\bibitem [{\citenamefont {Kepa}\ \emph {et~al.}(2016)\citenamefont {Kepa},
  \citenamefont {Ridley}, \citenamefont {Kamenev},\ and\ \citenamefont
  {Huxley}}]{kepapiston2016}%
  \BibitemOpen
  \bibfield  {author} {\bibinfo {author} {\bibfnamefont {M.~W.}\ \bibnamefont
  {Kepa}}, \bibinfo {author} {\bibfnamefont {C.~J.}\ \bibnamefont {Ridley}},
  \bibinfo {author} {\bibfnamefont {K.~V.}\ \bibnamefont {Kamenev}},\ and\
  \bibinfo {author} {\bibfnamefont {A.~D.}\ \bibnamefont {Huxley}},\ }\bibfield
   {title} {\bibinfo {title} {Piston cylinder cell for high pressure ultrasonic
  pulse echo measurements},\ }\href {https://doi.org/10.1063/1.4960082}
  {\bibfield  {journal} {\bibinfo  {journal} {Rev. Sci. Instr.}\ }\textbf
  {\bibinfo {volume} {87}},\ \bibinfo {pages} {085103} (\bibinfo {year}
  {2016})}\BibitemShut {NoStop}%
\bibitem [{\citenamefont {Eiling}\ and\ \citenamefont
  {Schilling}(1981)}]{eilingpressure1981}%
  \BibitemOpen
  \bibfield  {author} {\bibinfo {author} {\bibfnamefont {A.}~\bibnamefont
  {Eiling}}\ and\ \bibinfo {author} {\bibfnamefont {J.~S.}\ \bibnamefont
  {Schilling}},\ }\bibfield  {title} {\bibinfo {title} {Pressure and
  temperature dependence of electrical resistivity of {Pb} and {Sn} from
  1-{300K} and 0-10 {GPa}-use as continuous resistive pressure monitor accurate
  over wide temperature range; superconductivity under pressure in {Pb}, {Sn}
  and {In}},\ }\href {https://doi.org/10.1088/0305-4608/11/3/010} {\bibfield
  {journal} {\bibinfo  {journal} {J. Phys. F: Met. Phys.}\ }\textbf {\bibinfo
  {volume} {11}},\ \bibinfo {pages} {623} (\bibinfo {year} {1981})}\BibitemShut
  {NoStop}%
\bibitem [{\citenamefont {Yokogawa}\ \emph {et~al.}(2007)\citenamefont
  {Yokogawa}, \citenamefont {Yoshino},\ and\ \citenamefont
  {Aoyama}}]{yokojap07}%
  \BibitemOpen
  \bibfield  {author} {\bibinfo {author} {\bibfnamefont {K.~M.}\ \bibnamefont
  {Yokogawa}}, \bibinfo {author} {\bibfnamefont {H.}~\bibnamefont {Yoshino}},\
  and\ \bibinfo {author} {\bibfnamefont {S.}~\bibnamefont {Aoyama}},\
  }\bibfield  {title} {\bibinfo {title} {Solidificatoin of high-pressure medium
  daphne 7373},\ }\href@noop {} {\bibfield  {journal} {\bibinfo  {journal}
  {Jpn. J. Appl. Phys.}\ }\textbf {\bibinfo {volume} {46}},\ \bibinfo {pages}
  {3636} (\bibinfo {year} {2007})}\BibitemShut {NoStop}%
\bibitem [{\citenamefont {Sears}\ \emph {et~al.}(2017)\citenamefont {Sears},
  \citenamefont {Zhao}, \citenamefont {Xu}, \citenamefont {Lynn},\ and\
  \citenamefont {Kim}}]{sea17}%
  \BibitemOpen
  \bibfield  {author} {\bibinfo {author} {\bibfnamefont {J.~A.}\ \bibnamefont
  {Sears}}, \bibinfo {author} {\bibfnamefont {Y.}~\bibnamefont {Zhao}},
  \bibinfo {author} {\bibfnamefont {Z.}~\bibnamefont {Xu}}, \bibinfo {author}
  {\bibfnamefont {J.~W.}\ \bibnamefont {Lynn}},\ and\ \bibinfo {author}
  {\bibfnamefont {Y.-J.}\ \bibnamefont {Kim}},\ }\bibfield  {title} {\bibinfo
  {title} {Phase diagram of
  $\ensuremath{\alpha}\ensuremath{-}{\mathrm{rucl}}_{3}$ in an in-plane
  magnetic field},\ }\href {https://doi.org/10.1103/PhysRevB.95.180411}
  {\bibfield  {journal} {\bibinfo  {journal} {Phys. Rev. B}\ }\textbf {\bibinfo
  {volume} {95}},\ \bibinfo {pages} {180411} (\bibinfo {year}
  {2017})}\BibitemShut {NoStop}%
\bibitem [{Note1()}]{Note1}%
  \BibitemOpen
  \bibinfo {note} {We note that the sample size undergoes a change under
  applied pressure. At room temperature, we verified that the relative length
  change within the hexagonal plane remains below 3\% ($\Delta l/l_0 < 3\%$)
  for pressures up to 1 GPa. Based on XRD analysis \cite {bas18}, the
  sample-size change under pressure may be larger at lower temperatures. Since
  we do not have precise values, we did not include this correction in our
  analysis. We acknowledge that the reported sound velocity and its relative
  change might be slightly overestimated.}\BibitemShut {Stop}%
\bibitem [{\citenamefont {Kaib}\ \emph
  {et~al.}(2021{\natexlab{b}})\citenamefont {Kaib}, \citenamefont {Biswas},
  \citenamefont {Riedl}, \citenamefont {Winter},\ and\ \citenamefont
  {Valent\'{\i}}}]{Kaib2021}%
  \BibitemOpen
  \bibfield  {author} {\bibinfo {author} {\bibfnamefont {D.~A.~S.}\
  \bibnamefont {Kaib}}, \bibinfo {author} {\bibfnamefont {S.}~\bibnamefont
  {Biswas}}, \bibinfo {author} {\bibfnamefont {K.}~\bibnamefont {Riedl}},
  \bibinfo {author} {\bibfnamefont {S.~M.}\ \bibnamefont {Winter}},\ and\
  \bibinfo {author} {\bibfnamefont {R.}~\bibnamefont {Valent\'{\i}}},\
  }\bibfield  {title} {\bibinfo {title} {Magnetoelastic coupling and effects of
  uniaxial strain in $\ensuremath{\alpha}\ensuremath{-}{\mathrm{rucl}}_{3}$
  from first principles},\ }\href
  {https://doi.org/10.1103/PhysRevB.103.L140402} {\bibfield  {journal}
  {\bibinfo  {journal} {Phys. Rev. B}\ }\textbf {\bibinfo {volume} {103}},\
  \bibinfo {pages} {L140402} (\bibinfo {year}
  {2021}{\natexlab{b}})}\BibitemShut {NoStop}%
\bibitem [{\citenamefont {Yadav}\ \emph {et~al.}(2018)\citenamefont {Yadav},
  \citenamefont {Rachel}, \citenamefont {Hozoi}, \citenamefont {van~den
  Brink},\ and\ \citenamefont {Jackeli}}]{Yadav2018}%
  \BibitemOpen
  \bibfield  {author} {\bibinfo {author} {\bibfnamefont {R.}~\bibnamefont
  {Yadav}}, \bibinfo {author} {\bibfnamefont {S.}~\bibnamefont {Rachel}},
  \bibinfo {author} {\bibfnamefont {L.}~\bibnamefont {Hozoi}}, \bibinfo
  {author} {\bibfnamefont {J.}~\bibnamefont {van~den Brink}},\ and\ \bibinfo
  {author} {\bibfnamefont {G.}~\bibnamefont {Jackeli}},\ }\bibfield  {title}
  {\bibinfo {title} {Strain- and pressure-tuned magnetic interactions in
  honeycomb kitaev materials},\ }\href
  {https://doi.org/10.1103/PhysRevB.98.121107} {\bibfield  {journal} {\bibinfo
  {journal} {Phys. Rev. B}\ }\textbf {\bibinfo {volume} {98}},\ \bibinfo
  {pages} {121107} (\bibinfo {year} {2018})}\BibitemShut {NoStop}%
\bibitem [{\citenamefont {Balz}\ \emph {et~al.}(2019)\citenamefont {Balz},
  \citenamefont {Lampen-Kelley}, \citenamefont {Banerjee}, \citenamefont {Yan},
  \citenamefont {Lu}, \citenamefont {Hu}, \citenamefont {Yadav}, \citenamefont
  {Takano}, \citenamefont {Liu}, \citenamefont {Tennant}, \citenamefont
  {Lumsden}, \citenamefont {Mandrus},\ and\ \citenamefont {Nagler}}]{bal19}%
  \BibitemOpen
  \bibfield  {author} {\bibinfo {author} {\bibfnamefont {C.}~\bibnamefont
  {Balz}}, \bibinfo {author} {\bibfnamefont {P.}~\bibnamefont {Lampen-Kelley}},
  \bibinfo {author} {\bibfnamefont {A.}~\bibnamefont {Banerjee}}, \bibinfo
  {author} {\bibfnamefont {J.}~\bibnamefont {Yan}}, \bibinfo {author}
  {\bibfnamefont {Z.}~\bibnamefont {Lu}}, \bibinfo {author} {\bibfnamefont
  {X.}~\bibnamefont {Hu}}, \bibinfo {author} {\bibfnamefont {S.~M.}\
  \bibnamefont {Yadav}}, \bibinfo {author} {\bibfnamefont {Y.}~\bibnamefont
  {Takano}}, \bibinfo {author} {\bibfnamefont {Y.}~\bibnamefont {Liu}},
  \bibinfo {author} {\bibfnamefont {D.~A.}\ \bibnamefont {Tennant}}, \bibinfo
  {author} {\bibfnamefont {M.~D.}\ \bibnamefont {Lumsden}}, \bibinfo {author}
  {\bibfnamefont {D.}~\bibnamefont {Mandrus}},\ and\ \bibinfo {author}
  {\bibfnamefont {S.~E.}\ \bibnamefont {Nagler}},\ }\bibfield  {title}
  {\bibinfo {title} {Finite field regime for a quantum spin liquid in
  $\ensuremath{\alpha}\text{\ensuremath{-}}{\mathrm{rucl}}_{3}$},\ }\href
  {https://doi.org/10.1103/PhysRevB.100.060405} {\bibfield  {journal} {\bibinfo
   {journal} {Phys. Rev. B}\ }\textbf {\bibinfo {volume} {100}},\ \bibinfo
  {pages} {060405} (\bibinfo {year} {2019})}\BibitemShut {NoStop}%
\bibitem [{\citenamefont {Li}\ \emph {et~al.}(2021{\natexlab{b}})\citenamefont
  {Li}, \citenamefont {Said}, \citenamefont {Yan}, \citenamefont {Mandrus},
  \citenamefont {Lee}, \citenamefont {Okamoto}, \citenamefont {Halász},\ and\
  \citenamefont {Miao}}]{lidivergence2021}%
  \BibitemOpen
  \bibfield  {author} {\bibinfo {author} {\bibfnamefont {H.}~\bibnamefont
  {Li}}, \bibinfo {author} {\bibfnamefont {A.}~\bibnamefont {Said}}, \bibinfo
  {author} {\bibfnamefont {J.~Q.}\ \bibnamefont {Yan}}, \bibinfo {author}
  {\bibfnamefont {D.~M.}\ \bibnamefont {Mandrus}}, \bibinfo {author}
  {\bibfnamefont {H.~N.}\ \bibnamefont {Lee}}, \bibinfo {author} {\bibfnamefont
  {S.}~\bibnamefont {Okamoto}}, \bibinfo {author} {\bibfnamefont {G.~B.}\
  \bibnamefont {Halász}},\ and\ \bibinfo {author} {\bibfnamefont
  {H.}~\bibnamefont {Miao}},\ }\bibfield  {title} {\bibinfo {title} {Divergence
  of majorana-phonon scattering in kitaev quantum spin liquid},\ }\href@noop {}
  {\bibfield  {journal} {\bibinfo  {journal} {arXiv:2112.02015}\ } (\bibinfo
  {year} {2021}{\natexlab{b}})}\BibitemShut {NoStop}%
\end{thebibliography}%

\end{document}